\documentclass[prb,aps,amsmath,amssymb,amsfonts,twocolumn,superscriptaddress]{revtex4-1}

\usepackage[colorlinks=true,citecolor=blue,urlcolor=blue,linkcolor=blue,hyperfigures=true]{hyperref}
\usepackage{graphicx}
\usepackage{amsmath,amsfonts}
\usepackage{xcolor}
\usepackage{color}
\usepackage{subcaption}

\newcommand{\up}{\uparrow}
\newcommand{\dn}{\downarrow}

\newcommand{\av}[1]{\ensuremath{\left\langle #1 \right\rangle}}
\newcommand{\avb}[1]{\ensuremath{\left\langle #1 \right\rangle_\ast}}
\newcommand{\abs}[1]{\ensuremath{\left| #1 \right|}}

\usepackage{tikz}
\usetikzlibrary{calc}
\usetikzlibrary{arrows}
\usetikzlibrary{patterns}
\usetikzlibrary{decorations.pathmorphing}
\usetikzlibrary{decorations.pathreplacing}
\usetikzlibrary{decorations.markings}

\begin{document}

\title{Bandwidth renormalization due to the intersite Coulomb interaction}

\author{Yann in 't Veld}
\affiliation{Radboud University, Institute for Molecules and Materials, Heyendaalseweg 135, NL-6525 AJ Nijmegen, The Netherlands}

\author{Malte Sch\"uler}
\affiliation{Institut f{\"u}r Theoretische Physik, Universit{\"a}t Bremen, Otto-Hahn-Allee 1, 28359 Bremen, Germany}
\affiliation{Bremen Center for Computational Materials Science, Universit{\"a}t Bremen, Am Fallturm 1a, 28359 Bremen, Germany}

\author{Tim Wehling}
\affiliation{Institut f{\"u}r Theoretische Physik, Universit{\"a}t Bremen, Otto-Hahn-Allee 1, 28359 Bremen, Germany}
\affiliation{Bremen Center for Computational Materials Science, Universit{\"a}t Bremen, Am Fallturm 1a, 28359 Bremen, Germany}

\author{Mikhail I. Katsnelson}
\affiliation{Radboud University, Institute for Molecules and Materials, Heyendaalseweg 135, NL-6525 AJ Nijmegen, The Netherlands}

\author{Erik G. C. P. van Loon}
\email{evloon@itp.uni-bremen.de}
\affiliation{Radboud University, Institute for Molecules and Materials, Heyendaalseweg 135, NL-6525 AJ Nijmegen, The Netherlands}
\affiliation{Institut f{\"u}r Theoretische Physik, Universit{\"a}t Bremen, Otto-Hahn-Allee 1, 28359 Bremen, Germany}
\affiliation{Bremen Center for Computational Materials Science, Universit{\"a}t Bremen, Am Fallturm 1a, 28359 Bremen, Germany}

\begin{abstract}
The theory of correlated electrons is currently moving beyond the paradigmatic Hubbard $U$, towards the investigation of intersite Coulomb interactions.
Recent investigations have revealed that these interactions are relevant for the quantitative description of realistic materials. 
Physically, intersite interactions are responsible for two rather different effects: screening and bandwidth renormalization. 
We use a variational principle to disentangle the roles of these two processes and study how appropriate the recently proposed Fock treatment of intersite interactions is in correlated systems.
The magnitude of this effect in graphene is calculated based on cRPA values of the intersite interaction.
We also observe that the most interesting charge fluctuation phenomena actually occur at elevated temperatures, substantially higher than studied in previous investigations.
\end{abstract}

\maketitle

Simplicity of the Hamiltonian and complexity of the behaviour is the main characteristic of the Hubbard model. 
Although similar electronic models had existed for decades\cite{Schubin34}, the breakthrough of the 1960s~\cite{Hubbard63,Gutzwiller63,Kanamori63,Hubbard64} was to observe that only two parameters are sufficient to describe a rich spectrum of phenomena~\cite{Nagaoka66,Imada1998,Scalapino12} that occurs in correlated electron systems.
To go from a qualitative description of these phenomena to quantitative predictions for real materials, it is necessary to determine the parameters of the Hubbard model that correspond to any given material.   

The essential point is that comparatively simple methods can be used for the many weakly correlated bands, and to identify from this the appropriate correlated subspace with the corresponding Hubbard parameters. 
The state-of-the-art methods to do this are GW+DMFT~\cite{Biermann03} and cRPA~\cite{Aryasetiawan04}.
The former determines both the kinetic and the interaction terms of the Hubbard model explicitly in a Green's function approach. 
In the latter approach, the kinetic term is extracted from the DFT dispersion, whereas the interaction is determined using the cRPA approximation. 
The cRPA part could in the future be replaced by more sophisticated methods such as cFRG~\cite{Kinza15,Honerkamp18}.

In general, these methods provide an interaction which is not local. 
The bare Coulomb interaction is long-ranged.
Although screening by the weakly interacting bands reduces the effective range of the interaction, some spatial character generally remains, especially in two-dimensional compounds. 
The Hubbard model, however, contains only a local interaction and, importantly, the often used Dynamical Mean-Field Theory~\cite{Metzner89,Georges96,Kotliar06} for the solution of the Hubbard model is also restricted to local interactions. 
One way out is to extend the Hubbard model and to extend DMFT. 
This is under active investigation~\cite{Maier05,Rohringer18}, but it is complicated both conceptually and numerically. 
An alternative in the spirit of simple models is to try to capture the intersite interactions in the effective Hubbard parameters.

This approach was first studied in graphene, where it has been shown that screening due to the intersite interactions is crucial to keep graphene in its metallic state~\cite{Ulybyshev13,Schuler13}. 
It has been used further to study the effect of intersite interactions on the order of the metal-insulator transition~\cite{Schuler18}.
In these works, only the effective interaction was renormalized. 

In addition to screening the local interaction, another important physical effect of intersite interactions is the renormalization of the bandwidth~\cite{Ulstrup16}.
Recent work~\cite{Ayral17b} in the context of EDMFT~\cite{Sengupta95,Si96,Kajueter96,Smith00,Chitra00,Chitra01} has emphasized the importance of this effect. 
There, a perturbative treatment (the Fock diagram) of this effect was proposed. 
In strongly correlated systems, it is the question if such a perturbative treatment that uses Wick's theorem is valid and sufficient.
This question is particularly relevant now that non-equilibrium applications of EDMFT are appearing~\cite{Golez17,Golez18}.

In this work, we use a variational approach to study the bandwidth renormalization due to intersite interactions\footnote{A somewhat similar physical effect, the bandwidth renormalization due to phonons, has been studied~\cite{Casula12} using a Lang-Firsov transformation.}. 
The variational approach is not fundamentally limited to weakly correlated systems, so that we can use it to judge when a perturbative (Fock) treatment is reasonable. 
We compare bandwidth renormalization, interaction renormalization and a combined variational approach and explain how the physical regime determines which scheme is most effective.

To study the variational principle, we consider situations where the Hubbard model is exactly solvable: 
small systems where exact diagonalization is feasible and half-filled bipartite systems where QMC does not suffer from the sign problem. 
In the former, even the extended Hubbard model can be solved exactly, so that the applicability of the variational approach can be established. 
The latter scenario is directly relevant for graphene, which can be described as a half-filled single-band Hubbard model on the honeycomb lattice, and we use cRPA values for the intersite interaction to determine the bandwidth renormalization in graphene.

\section{Variational principle}

In this section, we provide a short overview of the variational principle, more details can be found in Appendix~\ref{app:derivation} and in Refs.~\onlinecite{Schuler13,vanLoon16b}. 
The idea of the variational principle is to describe an extended Hubbard model, $H$, with parameters $t$, $U$ and $V$ by an effective Hubbard model, $H^\ast$, with parameters $t^\ast$ and $U^\ast$, in formulas:
\begin{align}
 H =& - t^{\phantom{\ast}} \sum_{\av{i,j},\sigma} c^\dagger_{j\sigma} c^{\phantom{\dagger}}_{i\sigma} + U \sum_i n_{i\up} n_{i\dn} + \frac{1}{2} V \sum_{\av{i,j}} n_{i} n_{j} \notag \\
 H^\ast =& - t^\ast \sum_{\av{i,j},\sigma} c^\dagger_{j\sigma} c^{\phantom{\dagger}}_{i\sigma} + U^\ast \sum_i n_{i\up} n_{i\dn}.
\end{align}
Here $t$ is the hopping parameter, $U$ is the on-site Coulomb interaction, $V$ is the nearest-neighbor Coulomb interaction, $c^\dagger_{i\sigma}$ and $c^{\phantom{\dagger}}_{i\sigma}$ are creation and annihilation operators for an electron on site $i$ with spin $\sigma$ and $n_{i\sigma} = c^\dagger_{i\sigma} c^{\phantom{\dagger}}_{i\sigma}$ is the corresponding number operator. The total density on site $i$ is equal to $n_i=n_{i\up}+n_{i\dn}$. In the sum, $\av{i,j}$ denotes pairs of nearest-neighbor sites $i$, $j$ counted twice, i.e., both $ij$ and $ji$. 

We study this system either in the canonical ensemble (Sec.~\ref{sec:ed}) or in the grand-canonical ensemble at particle-hole symmetry (Sec.~\ref{sec:qmc}).
The latter corresponds to $\mu=U/2$ and an average density of one electron per site and $n_i$ is replaced by $n_i - 1/2$ in the intersite interaction term. 
For simplicity, we restrict ourselves to translationally invariant systems.

The true free energy of the extended Hubbard model is lower than the variational free energy~\cite{Peierls38,Bogoliubov58,Feynman72} of the effective Hubbard model for any choice of $t^\ast$ and $U^\ast$,
\begin{align}
 F \leq F_v=F^\ast + \avb{H-H^\ast}. 
\end{align}
Here $\avb{\cdot}$ denotes the expectation value of $\cdot$ with respect to the Hamiltonian $H^\ast$.
This inequality means that $t^\ast$ and $U^\ast$ can be used as variational parameters to minimize the variational free energy and to get an estimate of the true free energy $F$ of the original extended Hubbard model.
Furthermore, one might expect at least some observables of this effective Hubbard model to be close to the values of the original extended Hubbard model.

In previous works~\cite{Schuler13,vanLoon16b,Schuler18}, only a renormalization of the interaction strength $U^\ast$ was considered and the hopping parameter was kept at its original value $t^\ast=t$.
In this work, on the other hand, we focus exactly on the bandwidth renormalization. 
We will consider both bandwidth renormalization only, keeping $U^\ast=U$, which we will call the $t^\ast$ scheme, and the simultaneous variation of $t^\ast$ and $U^\ast$, which we call the ($U^\ast,t^\ast$) scheme.
In the former case, the bandwidth renormalization is determined by the minimum of the variational free energy and can be written as
\begin{align}
 t^\ast =& t + \alpha V \label{eq:tstar}, \\
 \alpha =& \frac{1}{4} \frac{\partial_{t^\ast} \avb{n_0 n_1}}{\partial_{t^\ast} G_{01}} \label{eq:alpha},
\end{align}
with $G_{01} = \avb{c^\dagger_{1,\sigma} c^{\phantom{\dagger}}_{0,\sigma}}$ the equal-time nearest-neighbor Green's function, which is proportional to the kinetic energy per site.
We note that in Eq.~\eqref{eq:tstar} $\alpha$ depends implicitly on $t^\ast$ since the expectation values are those of the effective Hubbard model.

An alternative to the variational principle is to use Hartree-Fock to determine the bandwidth renormalization. As shown in Appendix~\ref{app:HF}, this corresponds to
\begin{align}
 \alpha^{\text{HF}} = -G_{01}.
\end{align}
This result is derived from Eq.~\eqref{eq:alpha} by assuming that the effective Hubbard model $H^\ast$ is uncorrelated so that Wick's theorem can be applied. 
This assumption is not true for $U^\ast \neq 0$ and we will study how applicable the Hartree-Fock approximation is in Sec.~\ref{sec:hf}.

The variational principles give upper bounds for the free energy. 
Even when the exact free energy of the extended Hubbard model is not known, as is the case for the two-dimensional systems studied below, the methods can be compared since a lower upper bound is better. 
Variation over both $t^\ast$ and $U^\ast$ is obviously at least as good as varying over one of these parameters.

The variational principle fundamentally only gives an upper bound for the free energy of the system, without a direct way to see how good this bound is. 
For small values of $V$, the situation simplifies and some exact relations can be obtained by expanding the free energy as a power series in $V$. 
To linear order in $V$, the variational principle gives the exact free energy~\footnote{It is clear that there cannot be a linear term in $V$ in $F_{\text{true}}-F_\text{{variational}}$, since the quantity has to be positive for both signs of $V$. Since the two-parameter gives a free energy that is lower than the single-parameter schemes, it also gives the exact linear order in $V$ of the free energy. One could expect that the additional variational parameter also allows the exact second-order coefficient to be captured, but we will see that this is generallly not the case.}.
For the single parameter schemes, they also reproduce the exact value of their respective conjugate observables\cite{vanLoon16} to linear order in $V$, as shown in Appendix~\ref{app:exactobservables}.
The quadratic order in $V$ of the variational free energy is different in the two schemes, for our purposes it is sufficient to calculate how much free energy is gained by the variational principle, $\Delta F = F_v(U^\ast,t^\ast) - F_v(U,t)$, without needing to calculate the free energy of an extended Hubbard model. 
The result of this derivation, shown in Appendix~\ref{app:freeenergy}, is 
\begin{align}
 F_v(U,t^\ast) - F_v(U,t) =& - \frac{z V^2}{16} \frac{ \left(\partial_{t^\ast} \av{n_0 n_1}\right)^2}{\partial_{t^\ast} G} ,\notag \\
 F_v(U^\ast,t) - F_v(U,t) =& - \frac{z^2 V^2}{8} \frac{\left( \partial_{U^\ast} \av{n_0 n_1}\right)^2}{\partial_{U^\ast} D}.
\label{eq:deltaF}
\end{align}
Here, $D=\av{n_{0\up} n_{0\dn}}$ stands for the double occupancy, the probability of having two electrons at the same site. The sign of the difference $F_v(U,t^\ast) - F_v(U^\ast,t)$ tells us which variational parameter works best. 

\section{Solving the Hubbard model}

The variational treatment of $V$ requires an exact solution of the Hubbard model, the reference system. 
In this section we explain the methods we use for this purpose.

\subsection{Exact diagonalization}
\label{sec:ed}

If the number of lattice sites is small, the system can be solved exactly by diagonalizing the Hamiltonian. This can be done for the effective Hubbard model, but also for the original extended Hubbard model. 
In this way, it is possible to compare the free energy (and other observables) of the effective system to see how well the variational principle works.

Exact diagonalization is limited to finite and rather small systems. We consider a one-dimensional chain of six atoms (a ``benzene ring''). Here, this system is used simply as a toy model, the relation to actual benzene is discussed in Appendix~\ref{app:benzene}.
The total number of electrons is either 6 (half-filling) or 5 (17\% doping), and we use the canonical ensemble.

Since the ED is both exact and computationally light, we can calculate even the numerical derivatives with high accuracy. 

For the nearest-neighbor interaction $V$, we consider both positive and negative values. 
Positive values correspond to repulsive interaction and are most physically intuitive for the Coulomb interaction between electrons.  
We find the study of negative $V$ useful for the determination of linear coefficients. 
Attractive $V$ also serves as an interesting test case~\cite{vanLoon18b} since the physics of phase separation starts to become relevant.
We restrict ourselves to $\abs{V} < U$. At $V=U/2$, a non-uniform phase with alternating high and low density sites becomes favorable in terms of the potential energy. 
In an infinitly large system, this could lead to a phase transition to a charge-ordered phase~\cite{Vonsovsky79,Vonsovsky79b}. 
There are no phase transitions in finite systems, so the exact free energy is a smooth function. 
However, quite sharp features foreshadow phase transitions in the infinite volume limit. 
Approximate methods can (and do) result in discontinuities in finite systems.

\subsection{Determinant Quantum Monte Carlo}
\label{sec:qmc}

The Exact Diagonalization is restricted to rather small systems. 
To study larger systems, we consider bipartite (honeycomb and square) lattices at half-filling using Determinant Quantum Monte Carlo methods\cite{blankenbecler_monte_1981}. 
In this scenario the Hubbard model is free from the fermionic sign problem. 
We use the open source \textsc{quest} code\footnote{``QUantum Electron Simulation Toolbox'' \textsc{quest} 1.3.0 A. Tomas, C-C. Chang, Z-J. Bai, and R. Scalettar, (\url{http://quest.ucdavis.edu/})} to perform the simulations. 

Regarding lattice sizes, for the honeycomb lattice we used lattices up to $16\times 16$ unit cells (512 atoms) and comparing lattices with linear dimension $L=4,8,12,16$ we found relatively good convergence for lattices larger than $4\times 4$ unit cells (32 atoms). 
We used $8\times 8$ unit cells for all honeycomb lattice results shown here. 
A more detailed analysis of finite-size effects can be found in the appendix of Ref.~\onlinecite{Schuler18}.
The variational principle is fundamentally applicable both to finite and to infinite systems. 

The determination of the effective Hubbard parameters only needs local and nearest-neighbor observables, which converge quickly with lattice size, and in this work we mostly stay away from phase transitions where careful extrapolation is needed.

As opposed to the Exact Diagonalization results, the QMC is computationally quite expensive and all results suffer from numerical noise due to a finite amount of MC steps.
The simulations are generally more difficult, and the noise is worse, at low temperatures. 
The variational principle requires us to take numerical derivatives of noisy data, we use the Savitzky-Golay\cite{savitzky_smoothing_1964} filter to improve the stability of this numerical derivative.

For these sytems, we only perform calculations for the effective Hubbard model, determining the effective Hubbard parameters corresponding to a particular value of $V$. 
We cannot solve the extended Hubbard model exactly and thus we can also not compare the observables with the exact results. 
Therefore, we will focus on comparing the effective Hubbard parameters.
The renormalization of the interaction in the square and honeycomb lattice has been studied previously~\cite{Schuler13,vanLoon16b,Schuler18}, here we focus on the renormalization of the bandwidth and on the combined scheme. 

Since the determination of the effective $t^\ast$ requires a derivative only with respect to $t^\ast$, the honeycomb lattice simulations have been performed for constant $U$ and $\beta$ and varying $t^\ast$. 
At constant $\beta$, increasing $t^\ast$ amounts to increasing $\beta t^\ast$, i.e., lowering the temperature. 
In particular, this means that the Monte Carlo simulations become more difficult at large $t^\ast$, as is visible in the numerical noise.

\section{Results}

\subsection{Free energy}

\begin{figure}
5 electrons, 6 sites
\includegraphics{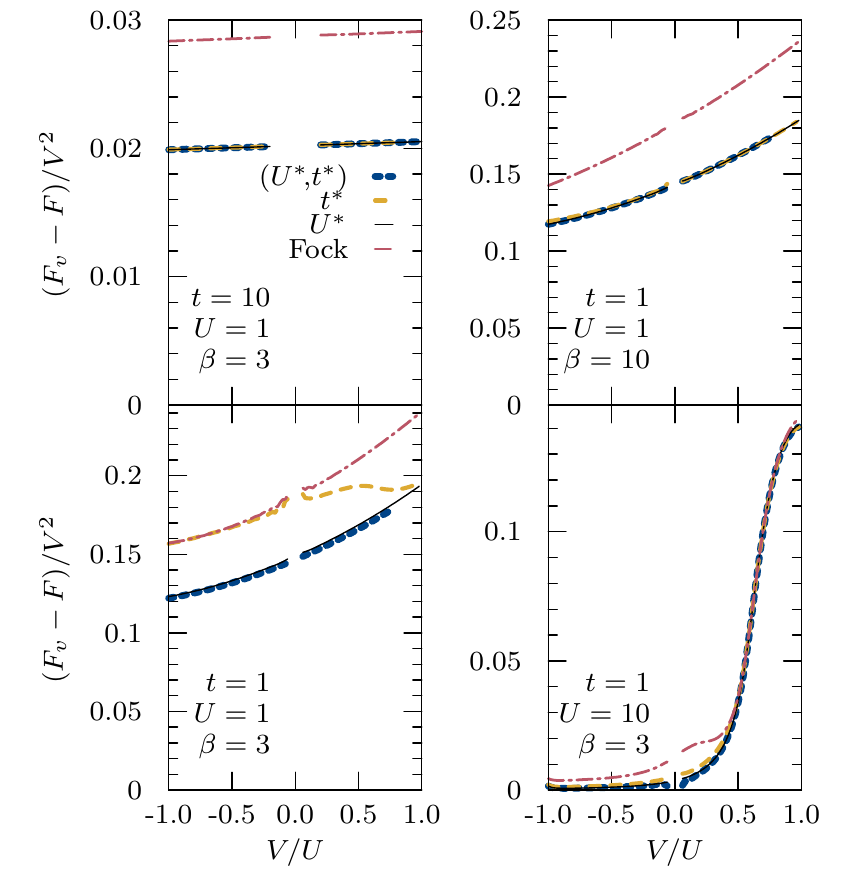}
\caption{Comparison of the variational free energy $F_v$ and the true free energy $F$ for a six-site chain with 5 electrons, solved using exact diagonalization. Results are only shown for $\abs{V}>0.05$ ($\abs{V}>0.5$ for the second panel). Figure~\ref{fig:energies:zoom} shows a zoomed in version of the last panel. Figure~\ref{fig:energies2} shows the same quantities at half-filling.
}
\label{fig:energies}
\end{figure}

\begin{figure}
\includegraphics{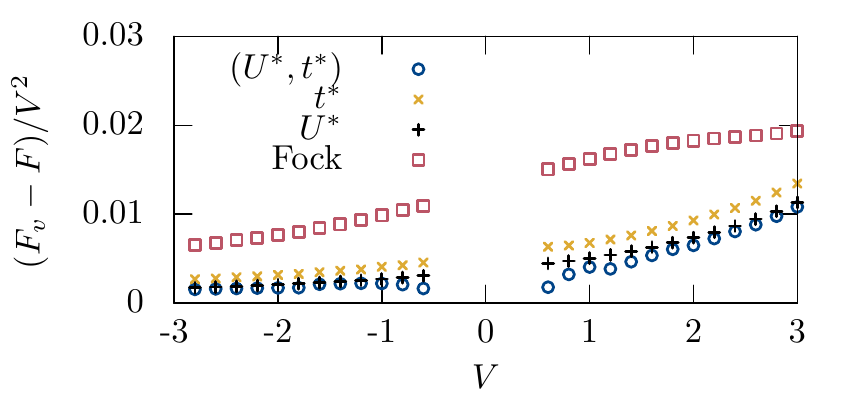}
\caption{Zoomed in version of Fig.~\ref{fig:energies}, corresponding to $t=1$, $U=10$, $\beta=3$, with 5 electrons in a six-site chain.}
 \label{fig:energies:zoom}
\end{figure}

We start our analysis with the free energy, since the variational principle is based on minimizing this quantity. We start our discussion with the variational schemes and come back to the Fock diagram in Sec.~\ref{sec:hf}. We only study the chain and the square lattice in this section.

\subsubsection{Chain}

Figure~\ref{fig:energies} shows the deviation between the variational free energy and the true free energy of the extended Hubbard model for a six-site chain with five electrons. To facilitate the comparison, following the discussion in Appendix~\ref{app:freeenergy} we have divided the free energy difference by $V^2$. 

Our first observation is that the variational free energy is indeed always larger than the true free energy of the system, which shows that the variational principle does not yield the exact solution of the extended Hubbard model. Furthermore, the variational free energy obtained from varying either of $t^\ast$ and $U^\ast$ is always larger or equal than the one obtained by varying both. For small $V$, the deviation from the true free energy is proportional to $V^2$ (all curves approach a finite number for $V\rightarrow 0$), showing that the variational principle does capture the linear term correctly.

At the lowest temperatures ($\beta t=10$ and $\beta t=3$), we see that there is hardly any difference between varying $t^\ast$, $U^\ast$ or both. In fact, as shown in Appendix~\ref{app:zeroT}, at $T=0$ these variational principles are completely equivalent since there is only a single physical parameter $U/t$. 
The numerical observations here show that this result also holds approximately at sufficiently low temperature.

For the first panel, where $U/t$ is very small, the lines are almost completely flat. 
This shows that the parabolic approximation of the variational free energy works very well in this regime.
For the second panel, deviations are visible albeit small.

The third panel shows a more elevated temperature, where the performance of the $t^\ast$ and $U^\ast$ schemes is different. For almost all values of $V$, the $U^\ast$ scheme gives a significantly lower free energy than varying $t^\ast$. There is little gain from varying both simultaneously.

Moving on to the fourth panel, $U/t=10$, the physics starts to be dominated by the potential energy. 
In particular, this means that there is a sharp crossover at $V=U/2=5$ when the alternating high and low density sites become favorable. 
A zoomed in version of the free energy at small $V$ is given in Fig.~\ref{fig:energies:zoom}. The figures shows that varying both $U^\ast$ and $t^\ast$ gives a better result than just varying either parameter in this situation.

The results of Fig.~\ref{fig:energies} correspond to a doped system. The situation in a half-filled system at these parameters is very similar and is given in Fig.~\ref{fig:energies2} in Appendix~\ref{app:freeenergy}.

\subsubsection{Square lattice}
\label{sec:free:square}

\begin{figure}
\includegraphics{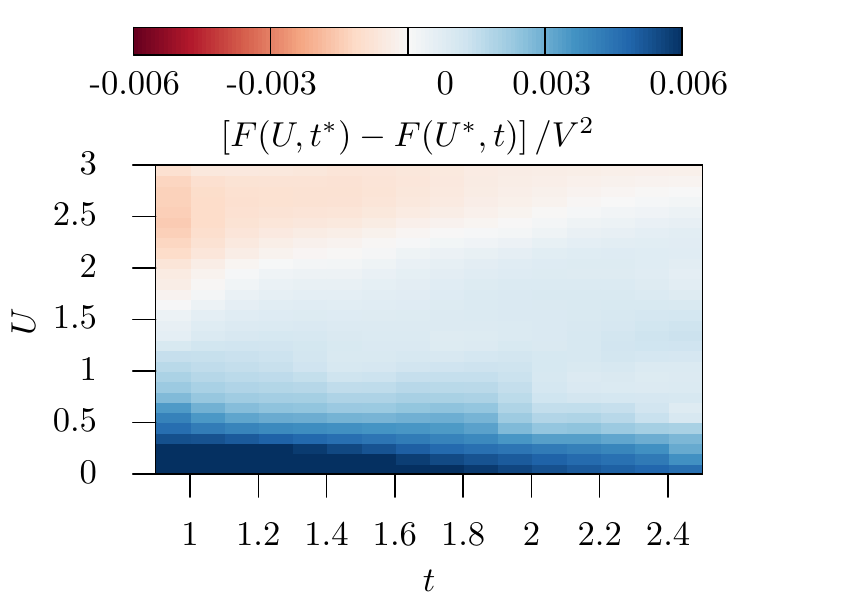}
\caption{Comparison of variational free energies obtained using the $t^\ast$ and $U^\ast$ optimization, for the square lattice model at $\beta=10$ and for infinitesimal $V$, see Eq.~\eqref{eq:deltaF}. The variation over $t^\ast$ performs better in the red region, the variation over $U^\ast$ in the blue region.}
 \label{fig:square:freeenergydifference}
\end{figure}

We briefly consider the square lattice, based on the simulations~\footnote{These differ from the other simulations in this work by the fact that finite-size extrapolation to the thermodynamic limit has been performed. This is of no concern to the current study since the variational principle is applicable both to finite and infinite systems.} of Ref.~\onlinecite{Schuler19}. 
Large scale scans over both $t^\ast$ and $U^\ast$ were performed, so that the two variational schemes can be compared. 
We use the expansion of the variational free energy for small $V$, as given in Eq.~\eqref{eq:deltaF} and Appendix~\ref{app:freeenergy}, to find out which variational method works best, bandwidth renormalization or interaction renormalization. 
Figure~\ref{fig:square:freeenergydifference} shows the difference in free energy between the $t^\ast$ and $U^\ast$ optimizations. 
We find that the bandwidth renormalization performs better at larger $U/t$ (red) and worse at small $U/t$ (blue). 

This observation might come as a surprise, but it can be understood in terms of energetics. When $U/t$ is large, both the extended and the reference Hubbard model have minimized their potential energy. The remaining low-energy degrees of freedom differ in their kinetic energy and this is captured well by $t^\ast$. 
When $U/t$ is small, the system minimizes its kinetic energy and the remaining low-energy degrees of freedom are governed by the potential energy, so the $U^\ast$ variational principle works better~\cite{Schuler19}. 
This is somewhat reminiscent of the Slater~\cite{Slater51} and Heisenberg~\cite{Anderson59} mechanisms of antiferromagnetism in the Hubbard model~\cite{Moukouri01,Kyung03,Gull08,Taranto12,Fratino17,vanLoon18}, with kinetic energy driven Heisenberg antiferromagnetism at large $U$ and potential energy driven Slater antiferromagnetism at small $U$.

\subsection{Effective parameters}

The variational principle minimizes the variational free energy by changing the effective parameters, so these are a natural second step in our investigation. 
The bandwidth renormalization corresponds to the value of $t^\ast$ that is found in the variational scheme.
The variational free energy landscape provides insight into the overall structure of the three variational schemes. We only show results for the chain model, where exact results are available.

\begin{figure}
\includegraphics[width=0.9\columnwidth]{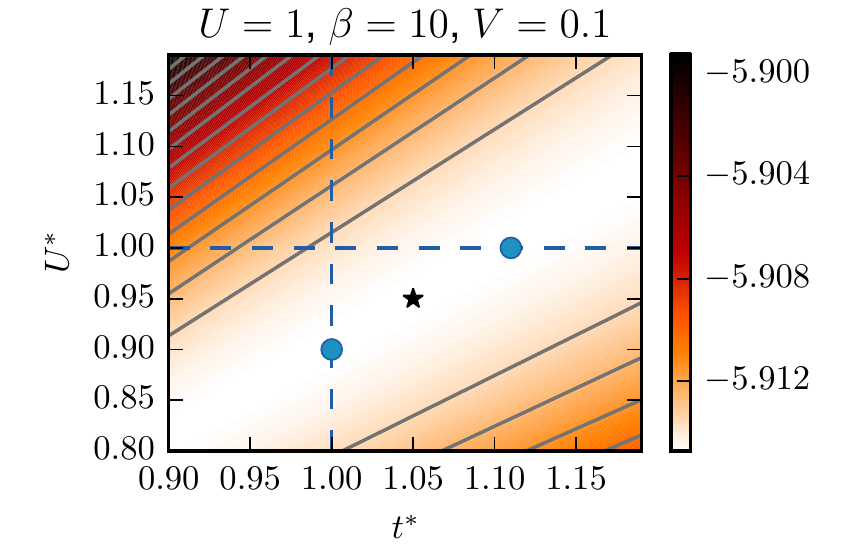}
\includegraphics[width=0.9\columnwidth]{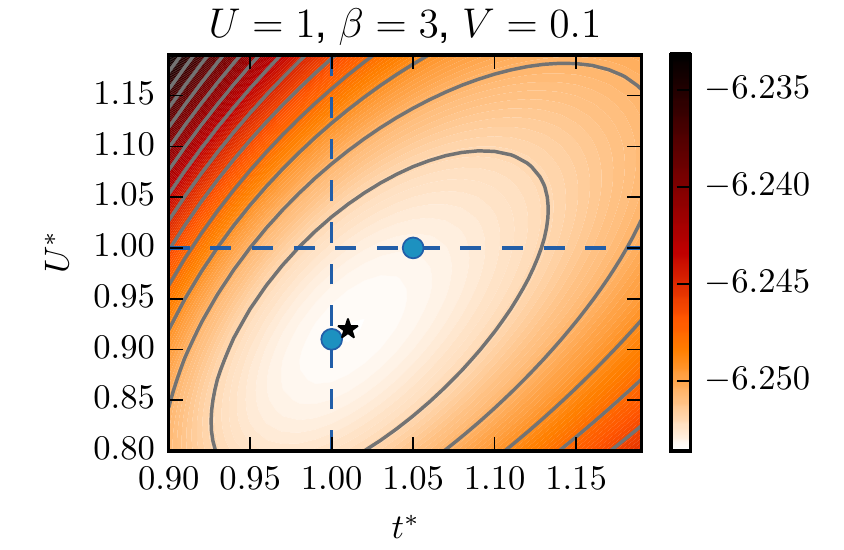}
\includegraphics[width=0.9\columnwidth]{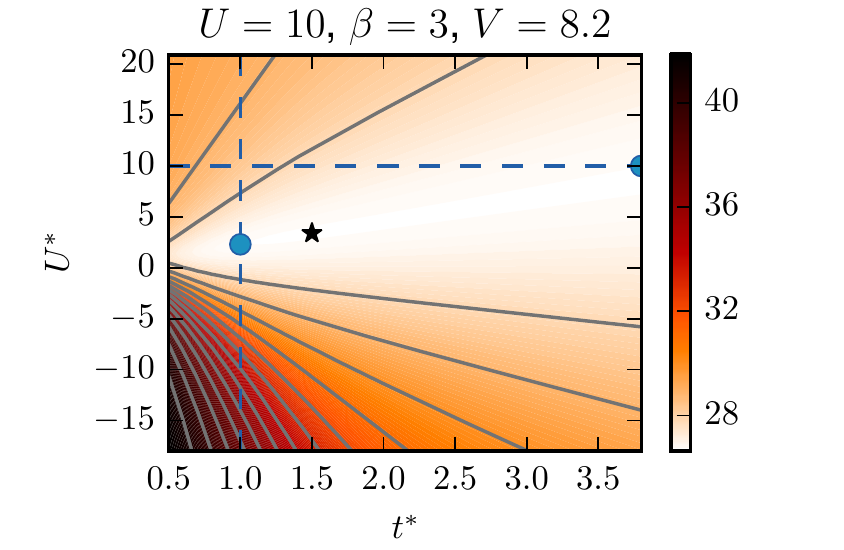}
\caption{Variational free energy surfaces for 5 electrons in a 6 atom chain. 
The color scale and the contour lines give the variational free energy functional $F_v(U^\ast,t^\ast)$. The star corresponds to the absolute minimum of the variational free energy, the two blue dots to the minima along the blue lines $t^\ast=t$ and $U^\ast=U$ corresponding to variational schemes with a single variational parameter.
Corresponding results at half-filling are given in Fig.~\ref{fig:freeenergysurface2}. All results have been obtained at $t=1$.
}
 \label{fig:freeenergysurface}
\end{figure}

Figures~\ref{fig:freeenergysurface} shows how the variational free energy depends on the effective parameters. At $\beta=10$, $U=1$ and $V=0.1$, the top panel, the free energy surface consists of curves of constant energy that are locally almost parallel lines. The $t^\ast$ and $U^\ast$ schemes are restricted to the dashed blue lines, with the associated minimum indicated by the blue dot. The global minimum found by the $(U^\ast,t^\ast)$-scheme is given by the black cross and is very close to the blue dot corresponding to the $U^\ast$ scheme. All renormalization schemes improve significantly on the original parameters $(U,t)$, i.e., on completely ignoring the intersite interaction. 

At $T=0$, the variational free energy landscape would be constant along lines of constant $U^\ast / t^\ast$. There would be one line with the lowest free energy and all three variational schemes would find the same free energy on this line. That the free energy landscape at $\beta=10$ still resembles the line picture of zero temperature supports the previous notion that it can be considered as almost zero temperature. 

At higher temperature $\beta=3$, shown in the second panel of Fig.~\ref{fig:freeenergysurface}, the free energy surfaces are ellipses. We find that the global minimum is quite close to $t^\ast=1$, so that changing only $U^\ast$ gives a better energy than changing only $t^\ast$. 

The third panel of Fig.~\ref{fig:freeenergysurface} shows a rather different situation, $U=10$ and $V/U=0.82>\frac{1}{2}$. Here, the nearest-neighbor repulsion is so strong that the electrons order themselves into alternating empty and doubly occupied sites~\cite{Vonsovsky79}. 
This ordering is not energetically favorable in the effective Hubbard model, so the variational principle cannot capture it properly.
This alternating pattern is even more favorable at half-filling, visible in Fig.~\ref{fig:freeenergysurface2}, where there are two local minima when varying $U^\ast$ along the line of constant $t^\ast=1$.
The previously seen discontinuities in the variational free energy correspond to the point where one of these minima overtakes the other as the global minimum. 
In this case, this first-order transition is a residual sign of the charge ordering. 
In fact, one of the two minima for $U^\ast$ and the global minimum for $(U^\ast,t^\ast)$ have an effective interaction that is attractive, $U^\ast<0$. 
Such an attractive interaction favors a high double occupancy.
This is the effective Hubbard model's way of describing the charge ordered phase, which indeed has many doubly occupied sites. 
What is lost in this effective local description is the alternating spatial character of the charge ordered state, which comes directly from the shape of the interaction $V$.

A comparison of Figs.~\ref{fig:freeenergysurface} and \ref{fig:freeenergysurface2}, which differ only in the filling, shows that only the charge ordering physics depends strongly on the density, the free energy surfaces for the other scenarios look very similar with 5 and with 6 electrons.

These parameters are rather extreme and not necessarily what one would expect in realistic scenarios (see Appendix~\ref{app:benzene}), we show them here to clearly illustrate how discontinuities coming from charge-density waves occur.
They are associated with a change of sign in the effective interaction $\tilde{U}$. This makes them look different from discontinuities associated with the metal-insulator transition~\cite{Schuler19}.

The free energy surfaces for the half-filled and the doped system look very similar for both $\beta=3$ and $\beta=10$, as can be seen by comparing Fig.~\ref{fig:freeenergysurface} and Fig.~\ref{fig:freeenergysurface2}.

Physically, what we see in these figures is that $V>0$ effectively makes the electrons more delocalized and more likely to doubly occupy a site. This is captured either by reducing the effective local interaction (lowering the potential energy penalty) or by increasing the hopping amplitude (increasing the potential energy gain from delocalization). Both correspond to reducing $U/t$, the balance between potential and kinetic energy in the Hubbard model.

\subsection{Applicability of Fock bandwidth renormalization}
\label{sec:hf}

Let us come back to one of our main questions, does Hartree-Fock describe the bandwidth renormalization due to $V$? 
The Hartree-Fock theory is perturbative in $V$ and can only be expected to work for small $V$, but even there the applicability depends on an approximation: using Wick's theorem for a correlated starting point.

\subsubsection{Chain}

For the six-site Hubbard chain, the results of Fock bandwidth renormalization are shown in Fig.~\ref{fig:energies}. 
We observe that the Fock result gives a free energy that is strictly larger than the variational approaches.
Only at $t=1$, $U=1$ and $\beta=3$ is the Fock result comparable with the $t^\ast$ scheme. 
This is also exactly in the regime where the Fock diagram could be expect to be reliable, $U/t$ is small and the temperature is high so that correlations should be moderate and Wick's theorem applies.
However, even though the Fock scheme gives considerably larger free energy deviations than the variational schemes, they are still clearly of the same order of magnitude, usually only 30\% larger and they also show the qualitative trends.
This indicates that, although the variational principle performs better, the Fock estimate is still useful as a poor man's approach to band width renormalization, especially in the moderately correlated regime. 

\subsubsection{Honeycomb lattice}

\begin{figure}
\includegraphics{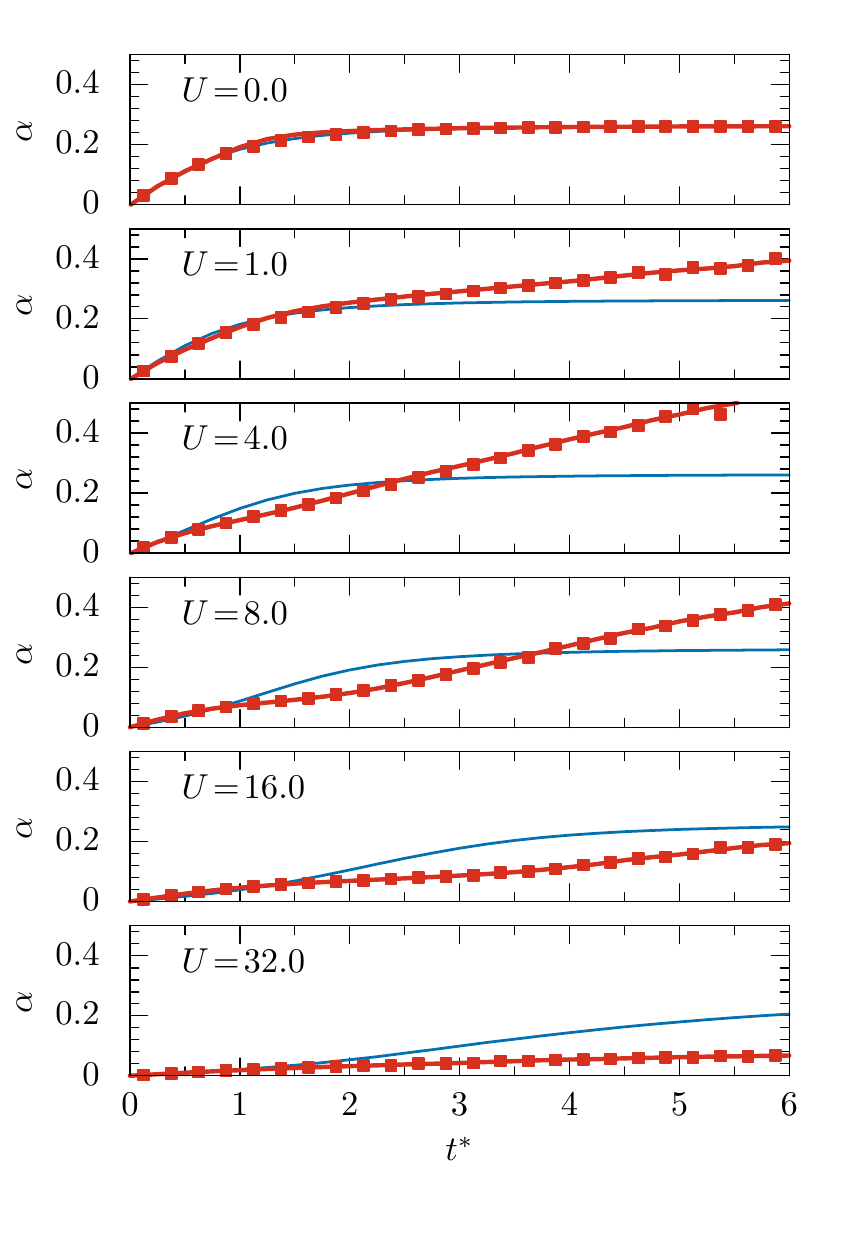}
\caption{Bandwidth renormalization $\alpha$, with $t^\ast = t+\alpha V$, according to Hartree-Fock (blue line) and according to the variational principle based on QMC results (red symbols). All results at fixed inverse temperature $\beta=1$. }
\label{fig:hexagon:alphaHF}
\end{figure}

Figure~\ref{fig:hexagon:alphaHF} shows the effective renormalization for various values of $U$ and $t^\ast$ (red symbols). 
As in the Exact Diagonalization results, we find that $\alpha>0$, i.e., the bandwidth gets wider due to intersite Coulomb repulsion $V>0$.
At small $t^\ast$, increasing $U$ leads to a monotonous decrease of $\alpha$. 
This can already be understood from the Hartree-Fock perspective, since $U$ reduces $G_{01}$.
At larger $t^\ast$, the situation changes and $\alpha$ is no longer a monotonous function of $U$.
For some values of $U$ ($U=0$, $U=32$), $\alpha$ seems to approach a constant for large $t^\ast$ whereas for intermediate values of $U$ the slope appears to stay non-zero within the studied range of $t^\ast$.

The figure also shows that as $t^\ast$ goes to 0, $\alpha$ also goes to 0. 
At $t=0$, the electrons cannot move and there is no dispersion, regardless of the value of $V$.
In particular, the nearest-neighbor Green's function is zero in the $t=0$ extended Hubbard model. 
We know that the variational principle must reproduce this, and the only way to do that is to keep $t=0$. 
The renormalization of the dispersion does not actually contain any information about $V$ in this extreme situation and the $t^\ast$ variational scheme is not applicable.

Figure~\ref{fig:hexagon:alphaHF} also compares the renormalization factor $\alpha$ obtained with the $t^\ast$ variational principle (red symbols) to the Hartree-Fock result (blue lines).
As expected, there is an exact match at $U=0$.
This happens because the Wick decoupling that is used to derive the Hartree-Fock expression is exact at $U=0$.
At small $U$, the Hartree-Fock results initially match the variational results but subsequently deviate at larger $t^\ast$.
Larger $t^\ast$ corresponds to larger $\beta t^\ast$, i.e., it effectively means that the temperature is lower.
Two-particle correlations are typically more important at low temperature, so that Hartree-Fock becomes less appropriate.
The results at $U=4$ and $U=8$ show that Hartree-Fock deviates from the variational results in both directions.
It initially overestimates the bandwidth renormalization and then saturates to a constant value of $\alpha$ that is too low.

Quantitatively, we observe deviations as large as a factor of 2 at intermediate and large interaction strengths.
This clearly shows the limits to the quantitative usefulness of the Hartree-Fock approach for incorporating bandwidth renormalization.
Whether Hartree-Fock overestimates or underestimates the bandwidth renormalization seems to depend quite sensitively on the value of $t^\ast$ and $U$.
However, in all cases studied, the Fock diagram at least produces the correct sign and order of magnitude of the bandwidth renormalization.

\subsection{Observables}

For the exact diagonalization results, we have access to all observables of the extended Hubbard model. 
This allows us to see how well the observables of the reference system match those of the true extended Hubbard model.
Theoretically, variational principles only make statements about (free) energies and there is no ground to identify observables of the reference system with those of the original system.
Practically, such an identification is still regularly made.

A previous study of the variational principle for the effective interaction~\cite{vanLoon16b} found two main conclusion regarding observables:
First, using the effective interaction leads to the \emph{exact} double occupancy at small $V$, since the double occupancy and the interaction strength are conjugate variables. 
Second, the variational principle does not predict the (momentum-resolved) charge susceptibility very well, since it depends explicitly on $V$ even in the weakly interacting limt. 
On the other hand, the prediction for many other observables is quite reasonable even though the variational principle technically only deals with the free energy. 

Regarding the first point, according to the same conjugate variable argument (see Appendix~\ref{app:exactobservables}), the $t^\ast$ scheme gives the exact value of the nearest-neighbor Green's function $\av{c^\dagger_0 c^{\phantom{\dagger}}_1}$, to linear order in $V$.
This exact statement is borne out by the numerical results shown in Appendix~\ref{app:exactobservables}.
In this case, the $U^\ast$ scheme actually only shows rather small deviations.

\begin{figure}
\includegraphics{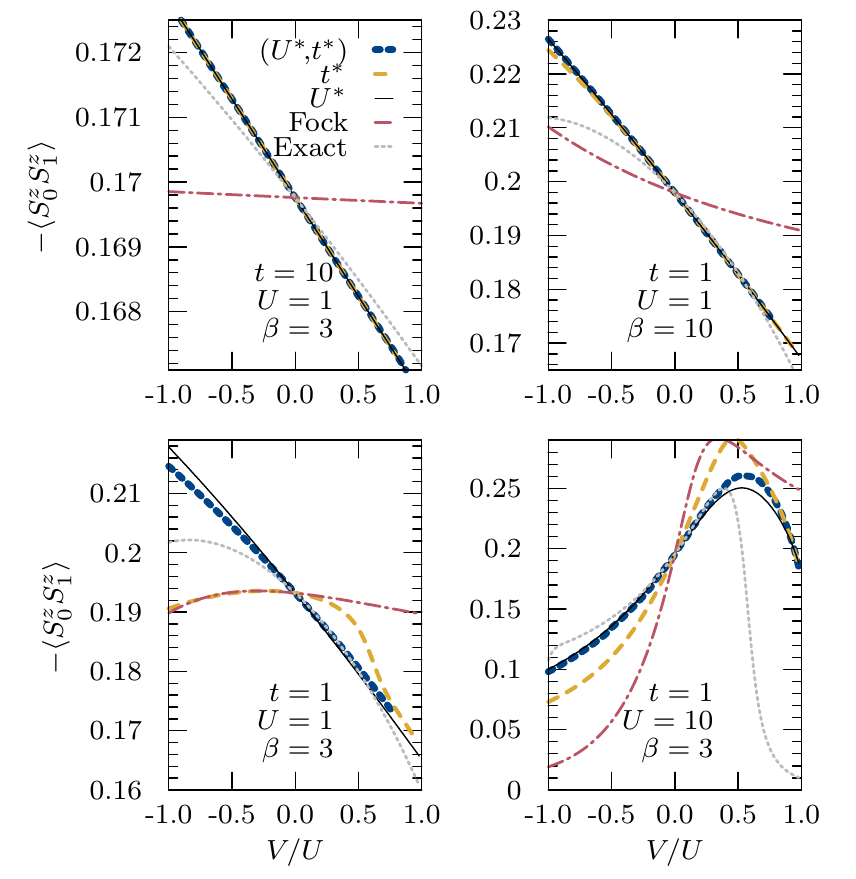}
 \caption{Nearest-neighbor spin correlation for a six-site chain with five electrons. The gray dashed line is the exact result for the extended Hubbard model, the other lines are the predictions of the variational approaches.
 }
 \label{fig:observables:Sz}
\end{figure}

For a generic observable not linked to any of the variational parameters, such as the nearest-neighbor spin correlation shown in Fig.~\ref{fig:observables:Sz}, none of the schemes necessarily predicts the correct linear coefficient in $V$. In fact, even in the moderately correlated regime at $t=10$, $U=1$, $\beta=3$ (top left), all variational schemes deviate from the exact solution already in linear order, although the deviations are not very large in an absolute sense.

Overall, the best predictions for the free energy were given by the $U^\ast$ and $(U^\ast,t^\ast)$ variational schemes and these also perform best at predicting the spin correlation. However, noticeable deviations from the exact result occur in all four panels, at large $V$. 

\begin{figure}
\includegraphics{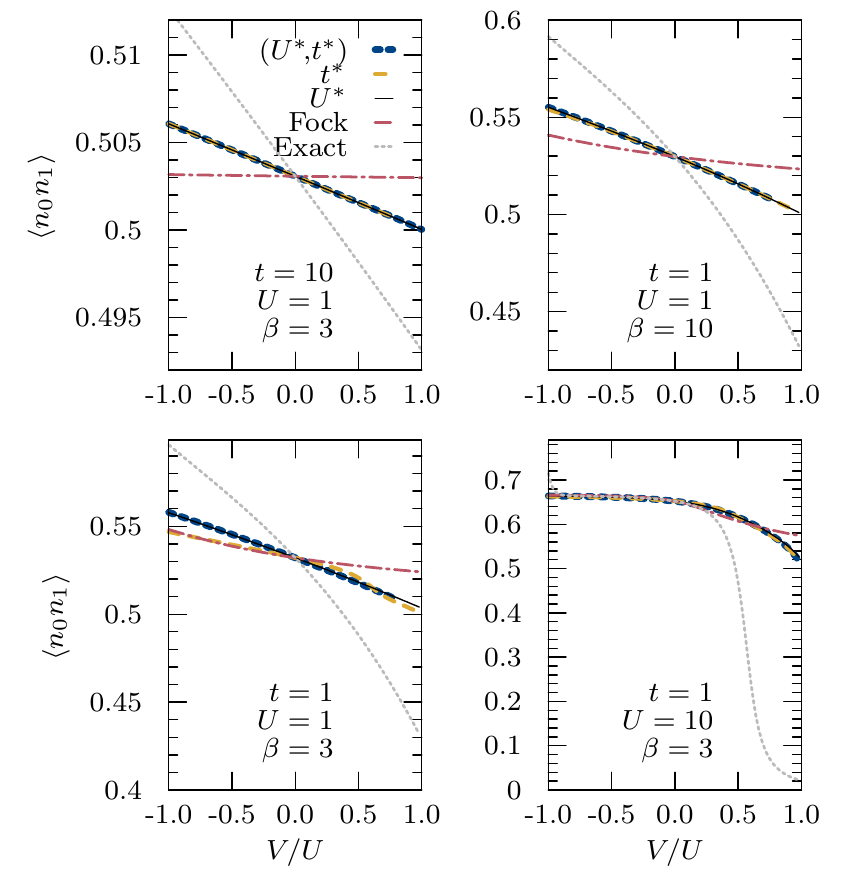}
\caption{Nearest-neighbor charge correlation for a six-site chain with five electrons, similar to Fig.~\ref{fig:observables:Sz}.
}
 \label{fig:observables:nn}
\end{figure}

Coming to the second point, the nearest-neighbor correlation function $\av{n_0 n_1}$ is a clear example of a quantity that depends explicitly on $V$. 
Indeed, we find that the variational principle does not capture this observable accurately, as shown in Fig.~\ref{fig:observables:nn}.
The variational principle underestimates how much the nearest-neighbor correlation function depends on $V$. 
This underestimation happens for all variational schemes and at almost all parameters we studied. 

\subsection{Graphene}

\begin{table}
\begin{center}
\begin{tabular}{c c c c c }
$\beta U$ & $\beta$ [eV$^{-1}$] & $t^*/t$ & $U/t^*$ & $\alpha$\\

\hline
$1.0$ & $0.1$ & $1.13$ & $3.33$ & $0.06$ \\ 
$4.0$ & $0.4$  & $1.30$ & $2.86$ & $0.14$ \\ 
$9.0$ & $0.9$  & $1.59$ & $2.22$ & $0.30$ \\ 
$16.0$ & $1.6$ & $2.07$ & $1.82$ & $0.51$ \\ 
\end{tabular}
\caption{Effective hopping parameters for graphene, using the parameters of Ref.~\onlinecite{Schuler13}, $t=2.8$ eV, $U/t=3.63$ and $V/t=2.03$.
}
\label{tab:graphene}
\end{center}
\end{table}

Table~\ref{tab:graphene} shows the bandwidth renormalization in graphene.
This is based on the graphene parameters of Ref.~\onlinecite{Schuler13}. 
As in the rest of this paper, we only consider the nearest-neighbor interaction $V$, with fixed $V/U = 2.03/3.63$. 

These results have been obtained from our simulations at $U=1$ and various inverse temperatures $\beta$. 
At fixed $\beta$ and $U$ $\alpha$, $\alpha$ is a function of $t^\ast$ only.
We then use the variational formula Eq.~\eqref{eq:alpha} to determine which $t^\ast$ corresponds to $t/U = 1/3.63$. 

Comparing the bandwidth renormalization with the interaction renormalization, here we find $U/t^\ast \approx 1.8$ for the bandwidth renormalization at the lowest temperature, which is comparable to the $U^\ast/t \approx 1.6 \pm 0.2$ obtained for interaction renormalization~\footnote{This result of Ref.~\onlinecite{Schuler13} includes interactions beyond nearest-neighbors, but for graphene the interaction renormalization based only on nearest-neighbors agrees within error bars, as can be seen in the supplemental material of Ref.~\onlinecite{Schuler13}.}.

Table~\ref{tab:graphene} shows significant temperature dependence, with the bandwidth renormalization getting stronger as the temperature is lowered. Here we should note that all temperatures listed here are at least one order of magnitude above room temperature when the graphene parameters are filled in, so that these results have to be extrapolated to make predictions about room temperature.
Given the temperature dependence observed, the bandwidth renormalization at room temperature is expected to be at least as strong as it is at the lowest temperatures studied here. We also note that the temperatures investigated here are already small compared to the intersite interaction, $\beta V \gg 1$.

The bandwidth renormalization comes from the intersite part of the Coulomb interaction $V$.
In an experimental set-up, this implies that control over the intersite interaction $V$ gives control over the effective bandwidth and the amount of correlation in the system.
For example, reducing $V$ should lead to a reduction of the bandwidth renormalization, so to a smaller bandwidth and a more correlated system.

\section{Conclusions and discussion}

Bandwidth renormalization is one of the most direct effects of intersite interactions. 
We have used the variational principle to study how the nearest-neighbor interaction in particular renormalizes the electronic bandwidth. 
We have found that the bandwidth of graphene can be widened by as much as 50\%-100\%. 

Determining the bandwidth renormalization is useful when a material is studied using a computational method that can only deal with the Hubbard model without intersite interactions, such as Dynamical Mean-Field Theory. 
The variational principle gives a good estimate for how much intersite interactions change properties of the system such as the free energy and the kinetic energy.
Of course, this estimate does not work very well for observables that depend explicitly on the spatial interactions, such as the charge-charge correlation function.
The variational principle also breaks down when the intersite interactions are so strong that they qualitatively change the physics of the system away from what can be in expected in a Hubbard model. 
The charge order physics visible in the third panel of Fig.~\ref{fig:freeenergysurface} is a good example of this.

Previously, it has been suggested~\cite{Ayral17} that a Hartree-Fock expression could be used to determine the bandwidth renormalization.
This approach assumes that the underlying Hubbard model is uncorrelated.
This assumption works well at small interaction strengths, where the system is only moderately correlated. 
On the other hand, it performs poorly at large $U$ or low temperature, as we have illustrated by comparing its results with the variational principle. 
Even then, qualitatively, it at least predicts the correct sign of the bandwidth renormalization, but more detailed aspects such as the non-monotonous dependence on $U$ (Fig.~\ref{fig:hexagon:alphaHF}) are not reproduced.

We have also identified some slightly pathological situations where the Hartree-Fock approach fails, mostly in few-electron systems where (spatial) correlations are naturally large, as illustrated in Appendix~\ref{app:1electron} and \ref{app:dimer}.  
Charge conservation induces these spatial two-particle correlations. Computational approaches for nanoscopic systems~\cite{Valli10,Valli15,Schuler17} need to take spatial correlations into account.

A perhaps somewhat surprising conclusion is that in our study, relatively high temperatures turn out to be more interesting than very low temperatures (and that in the benzene system, $\beta t =10$ is already low temperature).
The origin is that at higher temperatures, entropy competes with kinetic and potential energy and two relevant dimensionless parameters can be formed from $t$, $U$ and $T$.
At low temperature, only the ratio $U/t$ matters. 
The energy scale of charge fluctuations relevant in this study is given by $t$, $U$ and $V$ themselves, on the eV scale, instead of the much smaller emergent scales of spin and superconducting fluctuations in the Hubbard model.

From an experimental and computational point of view, there are advantages of studying charge fluctuations at temperatures comparable to the hopping. 
Recent experiments modelling the Hubbard model (without $V$) using ``ultracold'' fermions in optical lattices~\cite{Cocchi16,Cocchi17} were performed down to temperatures corresponding to $\beta t \approx 1.6$, similar to the values studied here.
Cluster approaches to the extended Hubbard model~\cite{Terletska17,Terletska18} are computationally lighter at high temperature and the phase diagrams of Ref. \onlinecite{Terletska18} go down to $\beta t \approx 5$.
For diagrammatic extensions of DMFT~\cite{Rohringer18}, at higher temperatures, the spin fluctuations are much less important allowing for a clearer vision on the charge fluctuations themselves and validating ladder approaches without feedback between the channels.

This suggests that future comparisons of computational approaches to the extended Hubbard model should occur not just at low temperatures but also at $T \approx t,U,V$.
In fact, the observation that similar phase boundaries~\cite{vanLoon14,Ayral17b} are found in a method based on bandwidth renormalization~\cite{Ayral17b} and those based on vertex corrections~\cite{vanLoon14}, more associated with interaction renormalization, might have to do with the fact that these comparisons were made at relatively low temperature ($\beta t=12.5$) where the Hubbard model has only a single parameter $U/t$. 
This single effective parameter might explain why two very different approaches ended up with the same numerical results.

\acknowledgments

The authors thank Alexander Lichtenstein for useful discussion. 
M.I.K. and E.G.C.P. v. L. acknowledge support from ERC Advanced Grant 338957 FEMTO/NANO.
The authors acknowledge the North-German Supercomputing Alliance (HLRN) for providing HPC resources under project \texttt{hbp00046}.

\appendix

\section{Derivation of variational approaches}
\label{app:derivation}

In this appendix, we give an overview of the derivations of the variational approaches dealing with the intersite interaction $V$. We restrict our analysis to nearest-neighbor interactions, since this situation is easiest to interpret as a renormalization of the bandwidth~\cite{Ayral17}. 
In general, $V$ can induce a change in electron density in addition to renormalization of the bandwidth and interaction. 
Within a variational framework, this can be described by using the chemical potential $\mu$ as an additional variational parameter. 
In this work, we only consider situations where this density change does not occur: finite systems in the canonical ensemble and a bipartite (graphene) lattice at particle-hole symmetry.  
Comparisons with experiment are also usually made by fixing the density.
The Hubbard model only provides a description of the low energy physics and going from band structure to Hubbard model involves shifts in the chemical potential anyway. 

\subsubsection{$t^\ast$}

The derivation of the variational formula for the bandwidth renormalization is similar to that of the effective interaction~\cite{Schuler13}. The Hamiltonians of the original and the effective system are
\begin{align}
 H =& - t^{\phantom{\ast}} \sum_{\av{i,j},\sigma} c^\dagger_{j\sigma} c^{\phantom{\dagger}}_{i\sigma} + U \sum_i n_{i\up} n_{i\dn} + \frac{1}{2} V \sum_{\av{i,j}} n_{i} n_{j} \notag \\
 H^\ast =& - t^\ast \sum_{\av{i,j},\sigma} c^\dagger_{j\sigma} c^{\phantom{\dagger}}_{i\sigma} + U \sum_i n_{i\up} n_{i\dn}.
\end{align}
To simplify the notation, we introduce the bandwidth renormalization as $\Delta t=t^\ast-t$, and calculate the required difference of the Hamiltonians.
\begin{align}
\frac{1}{N}\avb{H-H^\ast} =& \Delta t \sum_{\av{i,j},\sigma} \avb{c^\dagger_{j\sigma} c^{\phantom{\dagger}}_{i\sigma}} +  \frac{1}{2} V \sum_{\av{i,j}} \avb{n_{i} n_{j} } \notag \\
=& -\Delta t \cdot 2z  G_{01} +  \frac{V}{2} \cdot z  \avb{n_{0} n_{1} }
\end{align}
Here $z$ is the coordination number of the lattice and $G_{01} = -\frac{1}{2} \avb{c^\dagger_{1,\up} c^{\phantom{\dagger}}_{0,\up} + c^\dagger_{1,\dn} c^{\phantom{\dagger}}_{0,\dn}}$ is the Green's function \emph{averaged} over spin.

We continue with the derivative with respect to $t^\ast$,
\begin{align}
 \frac{1}{N}\partial_{t^\ast} \avb{H-H^\ast} =& -2z G_{01} + \Delta t 2 z \partial_{t^\ast} G_{01} \notag \\ &+ \frac{V}{2} z \partial_{t^\ast} \avb{n_0 n_1},
\end{align}
and the derivative of the free energy of the reference system $F^\ast$ with respect to $t^\ast$,
\begin{align}
\frac{1}{N} \partial_{t^\ast} F^\ast =& - \frac{1}{N} \sum_{\av{i,j},\sigma} \avb{ c^{\dagger}_{j\sigma} c^{\phantom{\dagger}}_{i\sigma} } \notag \\
=& 2z G_{01}. 
\end{align}

Combining the two gives,
\begin{align}
\frac{1}{N} \partial_{t^\ast} (F^\ast+\avb{H-H^\ast}) = \Delta t \cdot 2z \partial_{t^\ast} G_{01} + \frac{V}{2}z \partial_{t^\ast} \avb{n_0 n_1}, \label{eq:derivFreeEnergy}
\end{align}
so that the minimum is found when
\begin{align}
\Delta t =& V \frac{\partial_{t^\ast} \avb{n_0 n_1} }{4\partial_{t^\ast} G_{01}}. \label{eq:alphaT}
\end{align}
The factor of 4 essentially comes from two factors of 2, namely the sum over spins for the Green's function and the fact that the hopping term is directional (every bond counts twice) and the intersite interaction is not.

\subsubsection{$(U^\ast,t^\ast)$}

The derivation proceeds in a somewhat similar way when both the hopping and the interaction are taken as variational parameters.

\begin{align}
 H =& - t^{\phantom{\ast}} \sum_{\av{i,j},\sigma} c^\dagger_{j\sigma} c^{\phantom{\dagger}}_{i\sigma} + U \sum_i n_{i\up} n_{i\dn} + \frac{1}{2} V \sum_{\av{i,j}} n_{i} n_{j} \notag \\
 H^\ast =& - t^\ast \sum_{\av{i,j},\sigma} c^\dagger_{j\sigma} c^{\phantom{\dagger}}_{i\sigma} + U^\ast \sum_i n_{i\up} n_{i\dn}.
\end{align}
We introduce $\Delta t=t^\ast-t$ and $\Delta U = U-U^\ast$ (note the different sign).
\begin{align}
\avb{H-H^\ast} 
=& -\Delta t \cdot 2z  G_{01} + \Delta U \cdot D  + \frac{V}{2} \cdot z  \avb{n_{0} n_{1} },
\end{align}
with $D=\avb{n_{i\up} n_{i\dn}}$, and
\begin{align}
 F_v = F^\ast + \avb{H-H^\ast}.
\end{align}

One could now proceed in the same way as before and calculate the derivatives of $F_v$ with respect to $U^\ast$ and $t^\ast$, obtain the coefficient $\alpha$ for both parameters and so on. 
In practice, for the two-parameter optimization, we did not explicitly calculate numerical derivatives of the variational free energy, instead we simply calculated all necessary observables on a grid in $t^\ast$-$U^\ast$ space and subsequently determine the minimum and location of the minimum $(t^\ast,U^\ast)$ of the variational free energy $F_v$ for any desired combination of original variables $(t,U,V)$. 

\section{Hartree-Fock}
\label{app:HF}

The central idea of this work is to write the Hamiltonian of the extended Hubbard model as the sum of the Hubbard Hamiltonian and the intersite interaction,
\begin{align}
H_{\text{ext.}} = H_{\text{Hub.}} + V H_{\text{nl}}. \label{eq:H0}
\end{align}
The idea of this decomposition is that the first term on the left-hand side is solvable, either exactly as in the situations studied in this work or approximately using DMFT in realistic materials, whereas the combination is not. Instead of the variational approach, it is possible to consider perturbation theory in $V$ to deal with $H_{\text{nl}}$. 

If the Hubbard Hamiltonian would have been a non-interacting system, Wick's theorem would apply and the Hartree and the Fock terms would be the lowest order (linear in $V$) contributions in this perturbation theory. 
The Hartree term renormalizes the chemical potential, the Fock term can be interpreted as a renormalization of the hopping~\cite{Ayral17},
\begin{align}
 \Delta t_{ij} = - V_{ij} G_{ij}. \label{eq:hf:ayral}
\end{align}
It is important to note here that $G_{ij}$ corresponds to the \emph{exact} Green's function of the interacting Hubbard model, which plays the role of $H_0$ in Eqn.~\eqref{eq:H0}.
Because the Hubbard Hamiltonian describes a correlated system, even the linear in $V$ term in perturbation theory involves additional, more complicated terms that contain higher-order correlation functions (see Appendix~\ref{app:diagrams}). 
Hartree-Fock only captures the correct linear in $V$ behaviour when Wick's theorem applies, i.e., for $U=0$. 

The variational principle reproduces the Hartree-Fock expression in the uncorrelated limit, which shows that the variational principle also has the exact linear in $V$ terms at $U=0$. 
As mentioned before, this limit corresponds to the Wick decomposition of correlation functions, which states that higher-order correlations are simply a combination of single-particle Green's functions. 
This allows us to simplify the variational formula. It is most convenient to use the notation of the nearest-neighbor interaction in terms of density \emph{fluctuations},
\begin{align}
\av{V H_{\text{nl}}} =& \frac{zV}{2} \left( \av{n_0 n_1} - \av{n_0} \av{n_1} \right) ,
\end{align}
where $z$ is the number of nearest-neighbors. The Wick decomposition of the correlation function gives
\begin{align}
\av{n_0 n_1}  \overset{\text{Wick}}{=}& G_{00}G_{11} - G_{01}G_{10} \label{eq:app:wick1} \\
\av{n_0 n_1} - \av{n_0}\av{n_1}  \overset{\text{Wick}}{=}& - G_{01}G_{10}
\end{align}
This final expression contains the square of the nearest-neighbor Green's function and taking the derivative is simple,
\begin{align}
\partial_{t^\ast}\left( \av{n_0 n_1}- \av{n_0}\av{n_1}\right) =& - 2G_{01} \partial_{t^\ast} G_{01}.
\end{align}
Inserting this into the variational expression, the derivative of the Green's function cancels and we are left with the Hartree-Fock formula, Eq.~\eqref{eq:hf:ayral}. Note that this cancellation does not work when $\partial_{t^\ast} G = 0$, as in Appendix~\ref{app:1electron}.

The first term in Eq.~\eqref{eq:app:wick1} corresponds to the Hartree shift in the chemical potential, which is relevant when the Hamiltonian is written in terms of the density instead of the density fluctuations,
\begin{align}
\delta \mu = V G_{00}
\end{align}

To summarize, the Hartree-Fock expression for the bandwidth renormalization corresponds to two assumptions: small $V$ so that first-order perturbation theory in $V$ is justified \emph{and} the Wick decomposition of expectation values of the associated Hubbard model without $V$. The variational principle does not have the second assumption, in the sense that it is applicable at arbitrary interaction strength $U$ and, for some observables, even exact to first order in $V$, as shown in Appendix \ref{app:exactobservables}.

\section{Diagrammatic interpretation of the variational formula}
\label{app:diagrams}

The previous Appendix shows how the (diagrammatic) Hartree-Fock expression arises from the variational formula by assuming Wick's theorem. 
A diagrammatic interpretation of the complete variational formula is also possible, as we show in this Appendix. 
In this case, the diagrammatic expressions involve vertex corrections and higher-order correlators.
These are typically not accessible in actual calculations, so that the diagrammatic expression provided in this Appendix are not meant for computational purposes. 
Because of this, and to simplify the notation, we will proceed with the equations symbolically and will drop all numerical prefactors (including temperature), spin labels and site indices/momenta.
Regarding the last point, these equations should be understood as matrix equations in real space.

As in the Hartree-Fock theory, we again interpret $\Delta t = \alpha V$ as a self-energy (contribution) $\Sigma^V$. 
This is the linear contribution in $V$ towards the self-energy, starting from an interacting starting point $t_0$, $U_0$. 
All expectation values, Green's functions and correlation functions given below are those of the interacting starting point. Since the diagrammatic interpretation depends on the smallness of $V$, $t^\ast\approx t$ and we drop all $\ast$ labels to simplify the notation further.

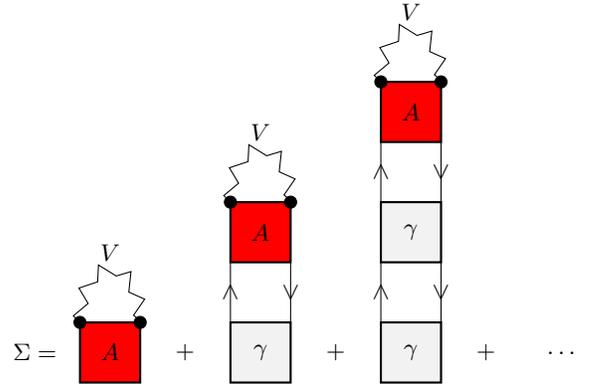
\begin{figure}
 \begin{tikzpicture}
  
  \begin{scope}[scale=0.8]
  
  \node at (-0.75,0.5) {$\Sigma=$} ;
  \node at (1.75,0.5) {$+$}  ;
  \node at (4.25,0.5) {$+$}  ;
  \node at (6.75,0.5) {$+$}  ;
  \node at (8.,0.5) {$\ldots$}  ;
  
  \draw[thick,fill=red] (0,0) -- (1,0) -- (1,1) -- (0,1) -- cycle ;
  \draw[fill=black] (1,1) circle (0.1) ;
  \draw[fill=black] (0,1) circle (0.1) ;
  \node at (0.5,0.5) {$A$} ;
  \draw[decorate,decoration=zigzag,out=90,in=90,looseness=3] (0,1) to node[above] {$V$} (1,1) ;

  \begin{scope}[shift={(2.5,2)}]
  
  \draw[thick,fill=gray!10] (0,-2) -- (1,-2) -- (1,-1) -- (0,-1) -- cycle ;
  \node at (0.5,-1.5) {$\gamma$} ;
  
  \draw[-] (0,-1) -- node[rotate=90] {$>$} (0,0) ;
  \draw[-] (1,-1) -- node[rotate=-90] {$>$} (1,0) ;
  
  \draw[thick,fill=red] (0,0) -- (1,0) -- (1,1) -- (0,1) -- cycle ;
  \draw[fill=black] (1,1) circle (0.1) ;
  \draw[fill=black] (0,1) circle (0.1) ;
  \node at (0.5,0.5) {$A$} ;
  \draw[decorate,decoration=zigzag,out=90,in=90,looseness=3] (0,1) to node[above] {$V$} (1,1) ;   
  \end{scope}

  \begin{scope}[shift={(5,4)}]

  \draw[thick,fill=gray!10] (0,-4) -- (1,-4) -- (1,-3) -- (0,-3) -- cycle ;
  \node at (0.5,-3.5) {$\gamma$} ;
  
  \draw[-] (0,-3) -- node[rotate=90] {$>$} (0,-2) ;
  \draw[-] (1,-3) -- node[rotate=-90] {$>$} (1,-2) ;

  \draw[thick,fill=gray!10] (0,-2) -- (1,-2) -- (1,-1) -- (0,-1) -- cycle ;
  \node at (0.5,-1.5) {$\gamma$} ;
  
  \draw[-] (0,-1) -- node[rotate=90] {$>$} (0,0) ;
  \draw[-] (1,-1) -- node[rotate=-90] {$>$} (1,0) ;
  
  \draw[thick,fill=red] (0,0) -- (1,0) -- (1,1) -- (0,1) -- cycle ;
  \draw[fill=black] (1,1) circle (0.1) ;
  \draw[fill=black] (0,1) circle (0.1) ;
  \node at (0.5,0.5) {$A$} ;
  \draw[decorate,decoration=zigzag,out=90,in=90,looseness=3] (0,1) to node[above] {$V$} (1,1) ;   
  \end{scope}
  
  \end{scope}
  
 \end{tikzpicture}

 \caption{Diagrammatic interpretation of the variational formula. The self-energy due to $V$ is given in terms of equal-time expectation values (vertices and propagators) of the $V=0$ Hubbard model.}
 \label{fig:app:diagram}
\end{figure}

Derivatives with respect to $t$ are equivalent to higher-order correlation functions.
The derivative in the denominator of Eq.~\eqref{eq:alpha} gives
\begin{align}
 \partial_t \av{c^\dagger c} =& \av{c^\dagger c c^\dagger c} - \av{c^\dagger c}\av{c^\dagger c} \notag \\
 =& GG\gamma GG - GG \notag \\
 =& -GG (1-\gamma GG),
\end{align}
where $\gamma$ denotes the two-particle vertex, a connected amputated two-particle correlation function\footnote{Here, the labels need to be treated with some care, since the two-particle vertex $\gamma$ is obtained from the correlation function by subtraction of two disconnected contributions. If we use labels $1,2,3,4$ to denote the momentum quantum number, for which the single-particle Green's function is diagonal, $\gamma$ is defined as
\begin{align*}
 G_1 G_2 G_3 G_4 \gamma_{1234} = \av{c^\dagger_1 c_2 c^\dagger_3 c_4}- G_1 G_3 (\delta_{12}\delta_{34}  - \delta_{14}\delta_{32}) 
\end{align*}
and we find
\begin{align*}
  \partial_{t_{12}} \av{c_3^\dagger c_4} =& \av{c^\dagger_1 c_2 c^\dagger_3 c_4} - \av{c^\dagger_1 c_2}\av{c^\dagger_3 c_4} \\
  =& G_1 G_2 G_3 G_4 \gamma_{1234}  - \delta_{14}\delta_{23}G_1 G_3 \\
  =& -G_1 G_3 (\hat{1}- G_{2}G_{4} \gamma_{1234} ),
\end{align*}
as given in the main text.
}.
The derivative in the numerator of Eq.~\eqref{eq:alpha}, $\partial_t \av{nn}$, is equal to a three-particle correlation function
\begin{align}
 \partial_t \av{nn} =& \av{n n c^\dagger c} - \av{nn}\av{c^\dagger c} \notag \\
 =& GG A,
\end{align}
which serves as the definition of the vertex $A$. 
This vertex has two amputated fermionic end points and two end points given by the density, which will couple to $V$.
Combining these elements into Eq.~\eqref{eq:alpha} gives
\begin{align}
 \Sigma^V = -\frac{AV}{1-GG\gamma},\label{eq:app:diagram}
\end{align}
where the denominator can now be interpreted as a geometric series of diagrams.
This is illustrated in Fig.~\ref{fig:app:diagram}.

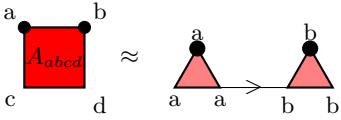
\begin{figure}
 \begin{tikzpicture}
  
  \begin{scope}[scale=0.8]
  
  \node at (1.75,0.5) {$\approx$};
  
  \draw[thick,fill=red] (0,0) -- (1,0) -- (1,1) -- (0,1) -- cycle ;
  \draw[fill=black] (1,1) circle (0.1) ;
  \draw[fill=black] (0,1) circle (0.1) ;
  \node at (0.5,0.5) {$A_{abcd}$} ;
  
  \node[anchor=north east] at (0,0) {c};
  \node[anchor=north west] at (1,0) {d};
  \node[anchor=south west] at (1,1) {b};
  \node[anchor=south east] at (0,1) {a};
    
  \end{scope}
  
  \begin{scope}[shift={(2,0)}]
  \draw[thick,fill=red!50] (0,0) -- (0:0.6) -- (60:0.6)-- cycle;
  \draw[fill=black] (60:0.6) circle (0.1) ;

  \node[anchor=north] at (0,0) {a};
  \node[anchor=south] at (60:0.6) {a};
  \node[anchor=north] at (00:0.6) {a};
  
  \end{scope}

  \begin{scope}[shift={(3.5,0)}]
  \draw[thick,fill=red!50] (0,0) -- (0:0.6) -- (60:0.6)-- cycle;
  \draw[fill=black] (60:0.6) circle (0.1) ;

  \node[anchor=north] at (0,0) {b};
  \node[anchor=south] at (60:0.6) {b};
  \node[anchor=north] at (00:0.6) {b};
  
  \end{scope}
  
  \draw[-] (2.6,0) -- node {$>$} (3.5,0) ;
  
 \end{tikzpicture}

 \caption{Under the assumption that correlations only occur locally, $A$ factorizes.}
 \label{fig:app:diagram:dmft}
\end{figure}

It is useful to think about the labels that occur in these diagrammatic expressions. 
Only equal time expectation values occur in the variational principle, so there is no frequency label in any of the vertices or Green's functions.
This aspect makes the theory much simpler than DMFT-based approaches which are based on the dynamic correlation functions of an auxiliary impurity model. 
On the other hand, the vertices here do have site labels. Looking at the vertex $A$, for example, there are four site labels occuring in two pairs, $A_{abcd} = \frac{\partial \av{n_a n_b}}{\partial t_{cd}}$. With both the interaction $V$ and the hopping $t$ restricted to nearest-neighbors, the labels of the vertex $A$ are two pairs of nearest-neighbors. 
In the DMFT spirit, we can then assume that the dominant correlations are between electrons on the same site, so that we only take the $\delta_{ac}\delta_{bd}$ contribution of $A_{abcd}$ into account. 
The second mean-field assumption is that $A_{abab}$ then factorizes into $\lambda_{a} G_{ab} \lambda_b$, where $\lambda$ denotes a local fermion-boson vertex (see Fig.~\ref{fig:app:diagram:dmft}.
This brings us back to the kind of diagrams regularly drawn in, for example, the dual boson theory~\cite{vanLoon14}). 

This construction suggests that if one is interested dealing with the bandwidth renormalization beyond Fock, in an impurity model based method, then the simplest approach might be to simply attach the equal-time fermion-boson vertex on both ends of the Fock diagram. For a single-band system, the equal-time impurity fermion-boson vertex is simply a number related to the double occupancy and the density.
Verifying such a simple scaling relation for the Fock energy could be done in a cluster approach to the extended Hubbard model~\cite{Terletska17,Terletska18}, since it has direct access to the self-energy and the double occupancy of the extended Hubbard model.
For a multiband system, the equal-time fermion-boson vertex is a matrix in orbital space. 

\section{Exact linear dependence of observables}
\label{app:exactobservables}

\begin{figure}
\includegraphics{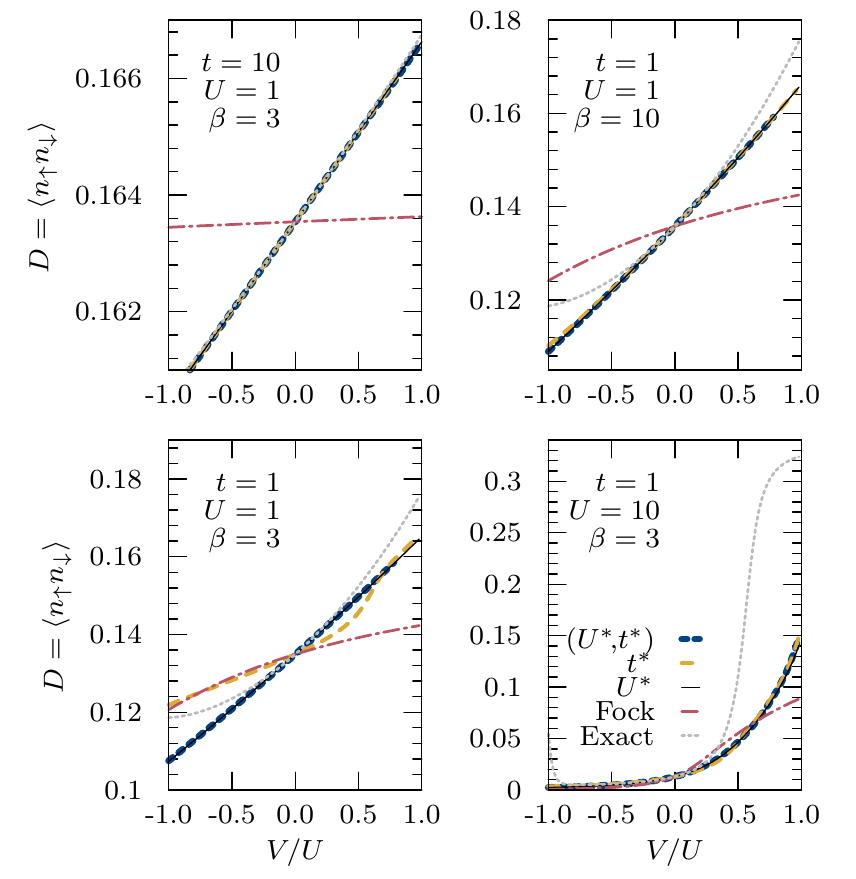}
 \caption{Double occupancy for a six-site chain with five electrons. The gray dashed line is the exact result for the extended Hubbard model, the other lines are the predictions of the variational approaches.}
 \label{fig:observables:D}
\end{figure}

\begin{figure}
\includegraphics{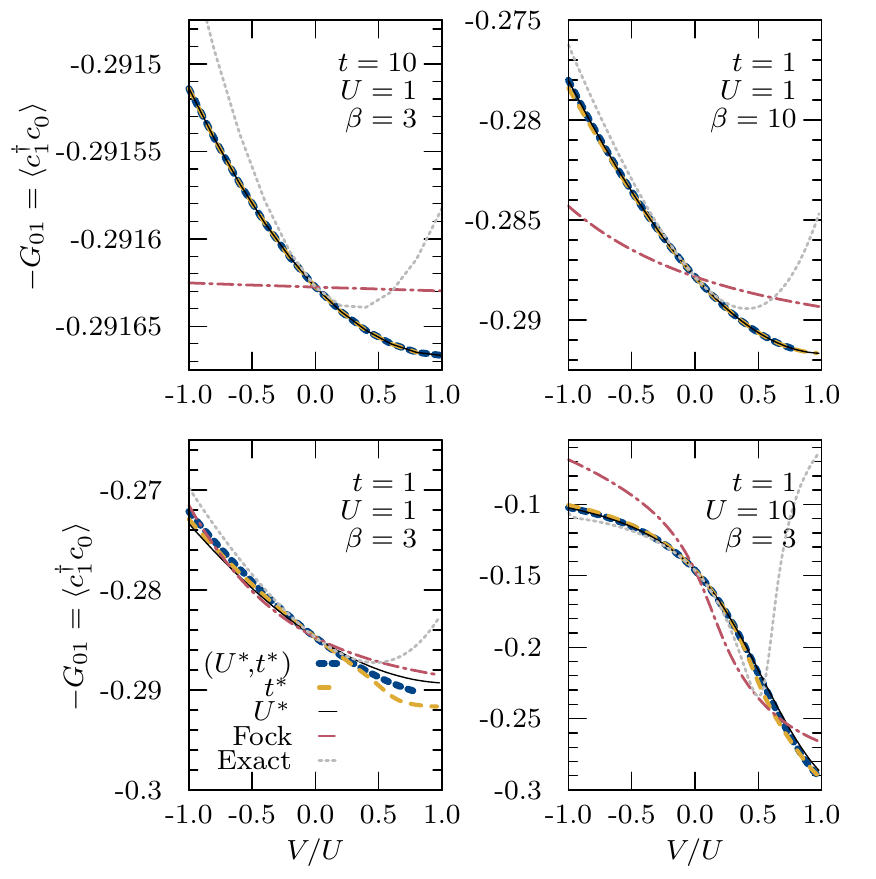}
 \caption{Nearest-neighbor Green's function for a six-site chain with five electrons. The gray dashed line is the exact result for the extended Hubbard model, the other lines are the predictions of the variational approaches.}
 \label{fig:observables:GF}
\end{figure}

The variational approach is non-perturbative and can in principle be appliead at arbitrary values of $V$. 
As usual with variational approaches, it is in general not guaranteed that the resulting observables correspond to the true value of the system.  
Formally, the variational approach just provides an upper bound for the free energy of the system.
However, at least at small $V$ it can be shown that the variational approach reproduces the exact linear in $V$ term of the observable conjugate to the variational parameter. 
This has been shown~\cite{vanLoon16b} for the double occupancy when the Hubbard interaction $U$ is the variational parameter.
A similar result holds for the hopping term when $t$ is the variational parameter.

The conjugate variable of $t$ is $-\av{c^\dagger_1 c^{\phantom{\dagger}}_0}=G_{01}$, the equal-time nearest-neighbor Green's function or the off-diagonal element of the density matrix. It is the first derivative of the free energy with respect to $t$, with a factor of two coming from the number of spin flavors,
\begin{align}
 G_{01} =& \frac{1}{2} \partial_t F.
\end{align}
This makes the derivative of the Green's function with respect to $V$ a second derivative of the free energy, 
\begin{align}
 \partial_V G_{01} =& \frac{1}{2} \frac{\partial^2 F}{\partial V \partial t} \\
 =& \frac{1}{4} \partial_t \av{n_0 n_1},
\end{align}
where the second line is obtained by interchanging the derivative operators.
If the Green's function depends smoothly on $V$, we find
\begin{align}
 G_{01}(t,V) \overset{\text{small $V$}}= G_{01}(t,V=0) + V \cdot \frac{1}{4} \partial_t \av{n_0 n_1} \label{eq:exactobservable1}
\end{align}
The variational principle predicts
\begin{align}
 G_{01}(t,V) &= G_{01}(t^\ast,V=0) \\
 &= G_{01}(t,V=0) + \partial_t G_{01} \cdot \Delta t \\
 &\overset{\eqref{eq:alphaT}}= G_{01}(t,V=0) + V \cdot \frac{1}{4} \partial_t \av{n_0 n_1}, \label{eq:exactobservable2}
\end{align}
and Eqs.~\eqref{eq:exactobservable1} and \eqref{eq:exactobservable2} are clearly identical. Note that this proof assumes the smallness of $V$, which implies smallness of $\Delta t$ so that it does not matter if the observables are calculated at $t$ or at $t^\ast$. 

This proof is based on the interchangeability of second derivatives and does not necessarily apply \emph{at} phase transitions, where the free energy is not smooth.

We illustrate these exact statements with results for the Extended Hubbard model with six sites. In particular, Fig.~\ref{fig:observables:D} shows that the $U^\ast$ scheme captures the exact linear coefficient in the double occupancy.
The $t^\ast$ and the Hartree-Fock schemes show significant deviations in the double occupancy at small $V$, as is visible in the $t=1$, $U=1$, $\beta=3$ results (bottom left). 
For the double occupancy, varying both $U^\ast$ and $t^\ast$ gives  results that are rather similar to varying just $U^\ast$.

For the $t^\ast$ scheme, the nearest-neighbor Green's function is the conjugate variable. Figure~\ref{fig:observables:GF} shows that this scheme indeed matches the exact nearest-neighbor Green's function to linear order in $V$. The $U^\ast$ scheme does not capture the linear coefficient exactly, but it is sufficiently close that the deviation is not visible in the figure. 

\section{Comparison of variational approaches}
\label{app:freeenergy}

With two variational parameters, $t^\ast$ and $U^\ast$, the question arises which variational parameter is best, that is, which parameter leads to the lowest free energy.
As in the proof of the exact observables of Appendix~\ref{app:exactobservables}, this question can be addressed for small $V$ simply by taking derivatives of the free energy. 

\begin{figure}
6 electrons, 6 sites
\includegraphics{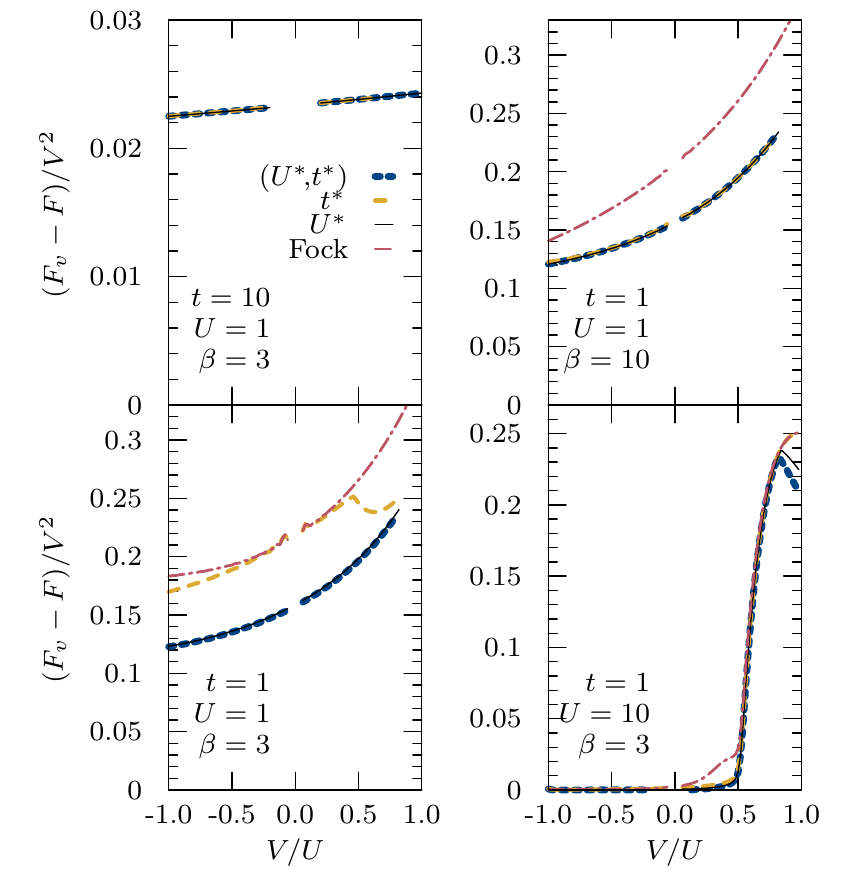}
\caption{
The same as Fig.~\ref{fig:energies}, but with 6 electrons on 6 sites (half-filling).
}
\label{fig:energies2}
\end{figure}

Let us start with the variation of the bandwidth. We know that there should be a minimum in the variational free energy at some $t^\ast$, and since $V$ is small we know that $t^\ast -t=\Delta t$ is small and that we can expand the variational free energy only up to second order in $t$ to find the location of the minimum.
To simplify the notation, we write the variational free energy functional around the original value $t_0$ as $\Phi(x)=F_v(t_0+x)/N$, in the following all derivatives are taken at $x=0$. A parabolic approximation of the free energy functional gives
\begin{align}
 \Phi(x) =& \Phi(0) + \partial_x \Phi \mid_{x=0} \cdot x + \frac{1}{2} \partial^2_x \Phi \mid_{x=0} \cdot x^2,
\end{align}
with minimum at
\begin{align}
 \Delta t = x_\text{min} = -\frac{\partial_x \Phi \mid_{x=0}}{\partial^2_x \Phi \mid_{x=0}}.
\end{align}
The value of the free energy functional at its minimum $x_\text{min}$ corresponds to the variational free energy.
\begin{align}
 \Phi(\Delta t) = \Phi(0) - \frac{1}{2} \frac{(\partial_x \Phi \mid_{x=0})^2}{\partial^2_x \Phi \mid_{x=0}} \label{eq:app:freeenergy:1}
\end{align}
Now, let us evaluate these derivatives as in Eq.~\eqref{eq:derivFreeEnergy}. 
\begin{align}
 \partial_x \Phi =& x\cdot 2z \partial_x G+ \frac{zV}{2} \partial_x \av{n_0 n_1} \\
 \partial^2_x \Phi =& x\cdot 2z \partial_x^2 G+2z \partial_x G +\frac{zV}{2} \partial^2_x \av{n_0 n_1}
\end{align}
We should note that we are evaluating these derivatives at the original value of $t$, i.e., at $x=0$ so that the first term in these equations vanishes,
\begin{align}
 \partial_x \Phi \mid_{x=0} =& \frac{zV}{2} \partial_x \av{n_0 n_1} \\
 \partial^2_x \Phi \mid_{x=0} =& 2z \partial_x G+\frac{zV}{2} \partial^2_x \av{n_0 n_1}.
\end{align}
The lowest (quadratic) order in $V$ is obtained by neglecting the second term in the second equation.
Going back to Eq.~\eqref{eq:app:freeenergy:1} and changing the derivatives back from $x$ to $t^\ast$, we obtain
\begin{align}
 \Phi(\Delta t) - \Phi(0) =& - \frac{z V^2}{16} \frac{ \left(\partial_{t^\ast} \av{n_0 n_1}\right)^2}{\partial_{t^\ast} G} .\\
\end{align}
The expansion in $U$ is very similar and gives
\begin{align}
 \Phi(\Delta U) - \Phi(0) =& - \frac{z^2 V^2}{8} \frac{\left( \partial_{U^\ast} \av{n_0 n_1}\right)^2}{\partial_{U^\ast} D}.
\end{align}
Together, we can determine which variational principle works best by looking at
$
\frac{1}{V^2}\left[\Phi(\Delta U) -\Phi(\Delta t)\right]
$, as is done in the main text.
For the exact diagonalization results, we have access to the true free energy and can use it as an absolute reference point, as is done in Fig.~\ref{fig:energies} and Fig.~\ref{fig:energies2}. 

\begin{figure}
\includegraphics[width=0.9\columnwidth]{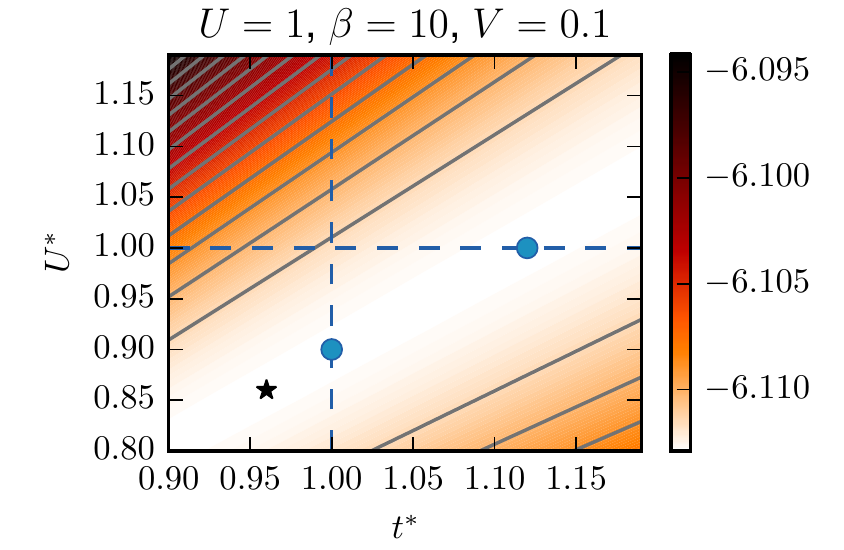}
\includegraphics[width=0.9\columnwidth]{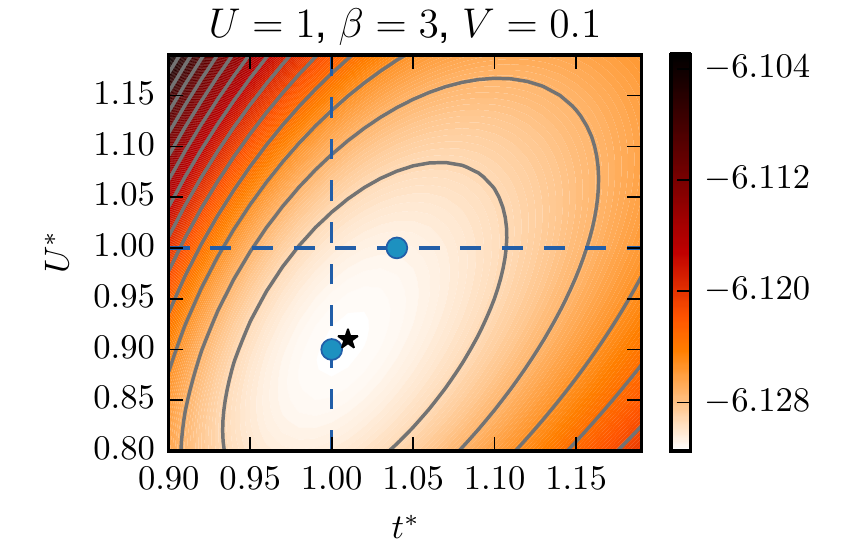}
\includegraphics[width=0.9\columnwidth]{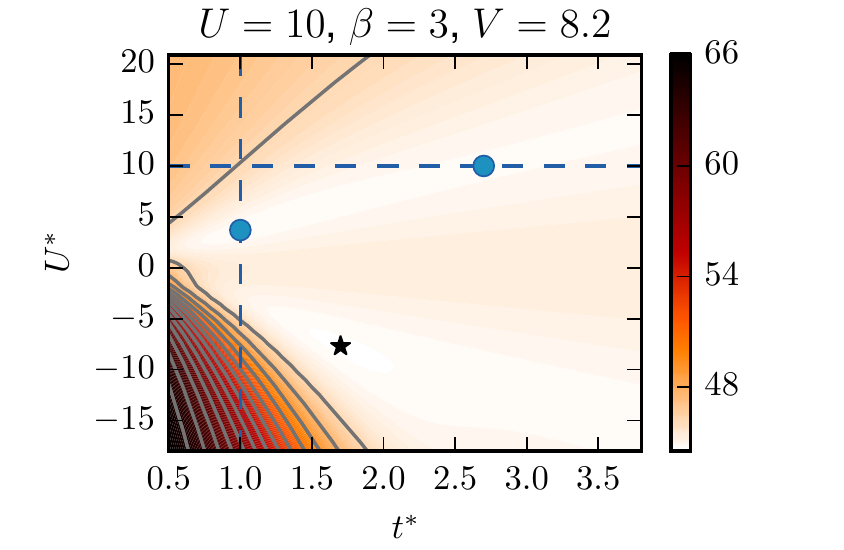}
\caption{Variational free energy surfaces for 6 electrons in a 6 atom chain. The same results for 5 electrons are given in Fig.~\ref{fig:freeenergysurface}. }
 \label{fig:freeenergysurface2}
\end{figure}

\section{Zero temperature}
\label{app:zeroT}

At zero temperature, there is effectively only a single parameter $U/t$ in the Hubbard model. This means that varying $t$ while keeping $U$ constant and vice versa results in the same set of density operators (ground states) $\rho_{U/t}$. 
Since both variational approaches have the same variational space, they find the same optimal density operator and corresponding observables. 
This shows that both variational approaches are equivalent at $T=0$.  
We expect this equivalence to still hold approximately at finite but very low temperature, if $T$ is smaller than any relevant effective energy scale in the problem. 

\section{Single electron}
\label{app:1electron}

An interesting situation to consider is a single electron in a finite lattice. Since there is only a single electron, there is no electron-electron interaction and both $U$ and $V$ do not change the properties of the system. In particular, this means that there is no renormalization of the bandwidth by $V$. 

To consider a single electron, we need to be careful about the statistical ensemble that is used. In the canonical ensemble, the particle number can indeed be fixed to a single electron. 
In the ensemble, we still average over configurations with one spin up electron and those with one spin down electron, so that the Green's function retains its spin symmetry. 
On the other hand, in the grand canonical ensemble at finite temperature, the ensemble averages always includes configurations with more than one electron and the bandwidth renormalization due to $V$ is finite.

The variational principle correctly captures the lack of self-interaction in the canonical ensemble. The numerator of the variational formula contains $\partial_{t*} \av{n_0 n_1}$, and $\av{n_0 n_1}=0$ since the single electron cannot be at site 0 and site 1 simultaneously. 
This results in $\Delta t =0$.

The diagrammatic derivation of the Hartree-Fock method at finite temperature naturally works in the grand canonical ensemble, where the density is not fixed and where the bandwidth renormalization is finite. 
Indeed, the Green's function in the Hartree-Fock expression does not vanish. 

At zero temperature, on the other hand, the density does not fluctuate and we are back to the situation where there should be no renormalization of the bandwidth. 
In fact, $t^\ast$ is the only dimensionful parameter in this case and as a result the dimensionless quantity $G_{01} \propto E_\text{kin}/t^\ast$ is independent of $t^\ast$. 
This means that both the numerator and the denominator of the variational formula are zero and that they cannot be divided out to derive the Hartree-Fock formula. 
So in this case, the Wick decoupling still works but does not lead to the Hartree-Fock expression.

Of course, it is good to note that although there is no electronic interaction in the single-electron system, there is correlation in some sense: $0=\av{n_{i,\up} n_{i,\dn}} \neq \av{n_{i,\up}} \av{n_{i,\dn}} = 1/(4N^2)$, with $N$ the number of lattice sites. 

\section{Hubbard dimer}
\label{app:dimer}

The Hubbard dimer, a system consisting of two Hubbard atoms, provides perhaps the simplest tractable example of intersite Coulomb interactions.
In this case, the variational mapping $U^\ast=U-V$ is exact~\cite{Schuler13}. 
The $t^\ast$ variational principle is not exact.

In the dimer, the Hartree-Fock approach performs poorly at any $U>0$. With only two sites in the system, any on-site correlation automatically corresponds to spatial correlation: if the second electron is not on site $1$ it is always on site $2$.

\section{Benzene}
\label{app:benzene}

\begin{figure}
\includegraphics[width=\columnwidth]{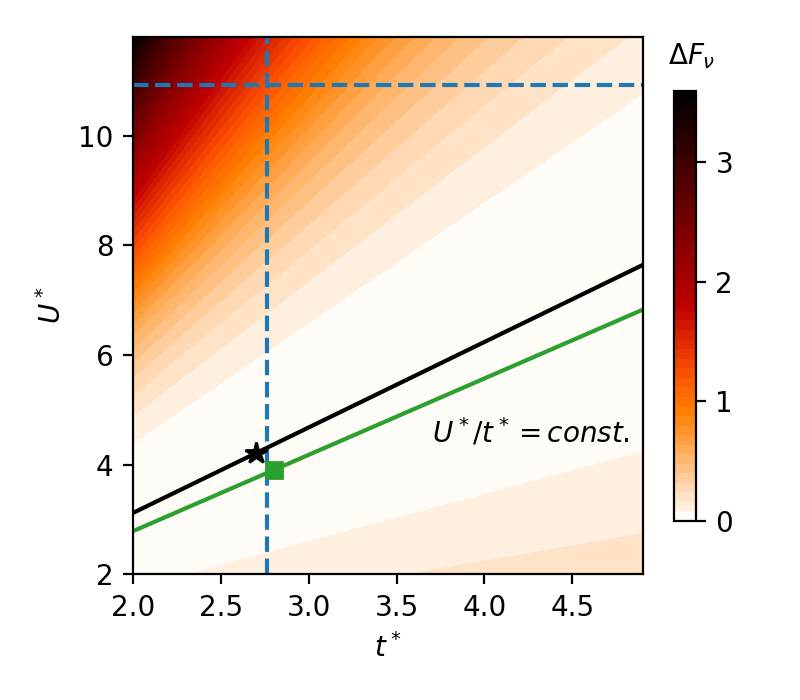}
\caption{Variational free energy for the benzene model. The black star corresponds to the optimal Hubbard model in our method, the green square to the optimal Hubbard model according to the DMC results of Ref.~\onlinecite{Changlani15}. The black and green lines denote constant $\tilde{U}/\tilde{t}$, which would be the only relevant parameter at zero temperature. The DMC values of the parameters are used, in particular $t=2.76$ and $U=10.92$ (blue dashed lines), all energies in units of eV. }
\label{fig:Wagner}
\end{figure}

We have used a periodic chain with six sites to illustrate the mapping of the extended Hubbard model onto an effective Hubbard model. A chain of six sites, with one orbital per site, is also a toy model for a benzene molecule~\cite{Pariser53,Pople53,Bursill98}. 
Ab-initio density matrix downfolding (AIDMD) has been used~\cite{Changlani15} to study the appropriateness of such a downfolding from 30 electrons (5 per site) to 6 (1 per site) in a single-orbital model. They present a downfolding both to a Hubbard model and to an extended Hubbard model, with the idea that both have a density matrix similar to the original benzene system.
Using Diffusion Monte Carlo (DMC), they find an extended Hubbard model with nearest-neighbor hopping $t=2.76$ eV, local interaction $U=10.92$ eV, nearest-neighbor interaction $V_{01}=7.13$ eV and next-nearest-neighbor interaction $V_{02}=5.41$ eV.  
In the same scheme, when downfolding to a Hubbard model with local interactions only, the optimal parameters are $t^*=2.80$ eV and $U^*=3.9$ eV, as illustrated by the green square in Fig.~\ref{fig:Wagner}. We have used the variational principle to map the extended Hubbard model for benzene onto an effective Hubbard model and find comparable values, as shown by the black star in Fig.~\ref{fig:Wagner}. The similar results of both techniques for mapping onto a Hubbard model, the variational principle and AIDMD, are a sign that these approaches give reasonable physical results. 

The main difference between the variational method and AIDMD is the temperature. The AIDMD is fundamentally based on zero temperature DMC calculations [although their scheme also involved solving the benzene molecule at a low temperature $T=1/20$ eV], whereas the variational principle is naturally formulated in terms of the free energy at finite temperature. In particular, according to the variational principle the parameters of the optimal Hubbard model can change as a function of temperature. For the benzene molecule, this temperature dependence is only relatively weak. In other situations, especially close to a metal-insulator transition, the temperature is more important \cite{Schuler19}.

Looking at observables, Changlani et al.\cite{Changlani15} also found that the Hubbard description is reasonable for many aspects except for the nearest-neighbor density correlation function. This conclusion is consistent with the results of Fig.~\ref{fig:observables:nn} and of Ref.~\onlinecite{vanLoon16b}.

The magnitude of the non-local Coulomb interactions in benzene is considerable compared to the on-site interaction. In fact, even though $V>U/2$, charge-ordering does not occur due to the large next-nearest-neighbor interaction. The competition between local and nonlocal interactions leads to a considerably smaller value of $U^\ast/t^\ast$ in the effective Hubbard model. In this example, the renormalization of the interaction $U^\ast$ is large and the bandwidth renormalization is relatively small. The predominance of interaction renormalization over bandwidth renormalization in the six site chain is consistent with our results in the main text.

\bibliography{references}
 
\end{document}